\documentclass[english,aip,jcp,numerical,reprint]{revtex4-1}
\usepackage{xcolor}
\usepackage[T1]{fontenc}
\usepackage{graphicx}
\usepackage{amsmath,amssymb,amsthm}
\usepackage{babel}
\newcommand{\bra}[1]{\ensuremath{\langle#1|}}
\newcommand{\ket}[1]{\ensuremath{|#1\rangle}}

\newcommand{\dg}{\dagger}
\newcommand{\tr}{\mathrm{tr}}

\begin{document}

\title{Dissipation enhanced vibrational sensing in an olfactory molecular switch}
\author{Agata Ch\k{e}ci\'{n}ska} 
\altaffiliation{These authors contributed equally to the work.}
%\email{agatachecinska@yahoo.com}
\affiliation{Centre for Quantum Technologies, National University of Singapore,
Singapore 117543}
\author{Felix A. Pollock}
\altaffiliation{These authors contributed equally to the work.}
%\email{felix.pollock@gmail.com}
\affiliation{Atomic \& Laser Physics, Clarendon Laboratory, University of Oxford, Parks Road,
Oxford OX1 3PU, United Kingdom}
\author{Libby Heaney}
\affiliation{Centre for Quantum Technologies, National University of Singapore,
Singapore 117543}
\author{Ahsan Nazir}
%\email{ahsan.nazir@manchester.ac.uk}
\affiliation{Photon Science Institute and School of Physics \& Astronomy, University of Manchester, Oxford Road, Manchester M13 9PL, United Kingdom}
\affiliation{Centre for Quantum Dynamics, Imperial College London, London SW7 2AZ, United Kingdom}
\date{\today}
\begin{abstract}
Motivated by a proposed olfactory mechanism 
based on a vibrationally-activated molecular switch, we study 
electron transport within a donor-acceptor pair that is 
coupled to a vibrational mode and embedded in a surrounding environment. We derive a polaron master equation with which we study the dynamics of both the electronic and vibrational degrees of freedom beyond previously employed semiclassical (Marcus-Jortner) rate analyses. We show:~(i) that in the absence of explicit dissipation of the vibrational mode, the semiclassical approach is generally unable to capture the dynamics predicted by our master equation due to both its assumption of one-way (exponential) electron transfer from donor to acceptor 
and its neglect of the spectral details of the environment; 
(ii) that by additionally allowing strong dissipation to act on the odorant vibrational mode we can recover exponential 
electron transfer, though typically at a rate that differs from that given by the Marcus-Jortner expression; 
(iii) that the ability of the molecular switch to discriminate between the presence and absence of the odorant, and its sensitivity to the odorant vibrational frequency, are enhanced significantly in this strong dissipation regime, when compared to the case without mode dissipation; and (iv) that details of the environment absent from previous %the earlier, 
Marcus-Jortner analyses can also dramatically alter the sensitivity of the molecular switch, in particular allowing its frequency resolution to be improved. 
Our results thus demonstrate the constructive role dissipation can play in facilitating sensitive and selective operation in molecular switch devices, as well as the inadequacy of 
semiclassical rate equations in analysing such behaviour over a wide range of parameters.  
\end{abstract}

\maketitle
\section{Introduction}
Characterising the influence of the environment on the transfer of charge and energy in an open quantum system is 
a problem of significant current interest.~\cite{nitzan_chemical_2006,may_kuhn_book,lambert_review13,cheng09,olaya_castro11,huelga11,breuer_theory_2002,ishizakireview10} 
In particular, resonant (or near resonant) interactions between environmental degrees of freedom and those inherent to the system are thought to play an important role in 
numerous physical processes.~\cite{garg85,olbrich11,shim12,phonon_antenna2013,chin_role_2013,oreilly_non-classicality_2014,Elinor_Irish2013,ritschel11,lim13,kolli12,christensson12,chenu13}  
However, a comprehensive 
picture of such dynamics is only beginning to emerge due to the complexity of the systems in question. 
Here, by focusing on a proposed model for olfaction as a vibrationally-activated molecular switch, we explore the detailed effects of the environment on the dynamics of electron transfer (ET) in an open quantum system, aiming to gain physical insight into vibrationally-assisted 
transport processes more generally.

In fact, obtaining a deeper understanding of olfaction, the mechanism for which is still being actively debated, \cite{turin_spectroscopic_1996,turin_method_2002,keller_psychophysical_2004,brookes_could_2007,hettinger_olfaction_2011,franco_molecular_2011,solovyov_vibrationally_2012,kovacic_mechanism_2012,bittner_quantum_2012,gane_molecular_2013} is an important problem in its own right for both fundamental science and industry.  \cite{rowe_chemistry_2005,axel_scents_2005,buck_unraveling_2005,lee_mimicking_2012,farahi_critical_2012} 
The prevailing theory, known as the lock-and-key model, explains how the odorant size and shape can provide discrimination in the receptor.~\cite{silverman_organic_2002} However, this theory does not give a straightforward explanation of why it may be possible to distinguish the scents of some very similar odorants, for example between those that are deuterated and non-deuterated.~\cite{haffenden_investigation_2001,franco_molecular_2011,gane_molecular_2013} It was suggested as early as 1938 that the sensing of vibrational spectra of molecules~\cite{dyson_scientific_1938,wright_odor_1977,wright_sense_1982}---later proposed to occur via electron transfer~\cite{turin_spectroscopic_1996,turin_method_2002}---could play an important role in olfaction, supplementing (rather than replacing) the existing lock-and-key model. Recent work, focussing on constructing and exploring model systems that capture the important physical processes,~\cite{brookes_could_2007,solovyov_vibrationally_2012} has shown that this is indeed a viable proposition. 
The suggested mechanism, which harnesses vibrationally-assisted ET in a similar manner to inelastic electron tunneling spectroscopy,~\cite{lambe_molecular_1968,adkins_inelastic_1985} can be viewed as an example of a molecular switch, wherein specific vibrations of an external molecule actuate the receptor and lead to a pronounced electron flow. Detection via the frequency of vibrations could help to discern two molecules that are otherwise very similar. 
However, the need for clear discrimination of the difference in ET dynamics in the presence and absence of the odorant 
imposes certain design principles on the molecular switch, which may or may not lead to the employment of quantum  phenomena to 
optimise performance. 

Previously, a receptor-odorant spin-boson model was introduced to describe the vibrationally-assisted ET process at the heart of the proposed mechanism.~\cite{brookes_could_2007,solovyov_vibrationally_2012} 
Using an analysis based on the semiclassical Marcus-Jortner (MJ) formula for the ET rate, \cite{marcus_theory_1956,marcus_chemical_1964,marcus_theory_1965,marcus_electron_1985,bixon_electron_1999} it was shown  that for certain parameter values the rates for ET in the presence and absence of the odorant can differ drastically.
This indicates how the ET 
process could help in molecular recognition via sensing of vibrational spectra; 
sensitivity of the switch to the presence or absence of the odorant can be understood as a significant difference in typical timescales, or more generally in the population dynamics, for processes with or without the odorant coupled to the receptor. 
Additionally, in Refs.~\onlinecite{solovyov_vibrationally_2012,bittner_quantum_2012} further evidence supporting a vibrationally-assisted mechanism was obtained from sophisticated quantum chemistry calculations. 

Inspired by the possibility of vibrational sensing in olfaction, 
we focus here on the physical model of a molecular switch tuned for frequency detection. We go beyond the MJ approach 
to look at both frequency selectivity and detailed dynamics in a variety of regimes in which a semiclassical analysis breaks down. Starting from a microscopic Hamiltonian describing the odorant (as an oscillator), receptor (as a donor-acceptor two-level system), and environment (as a collection of independent oscillators), we derive a polaron-representation master equation for the ET process.~\cite{holstein_studies_1959-1,holstein_studies_1959,JacksonSilbey1983,nazir09,jang_theory_2008,jang09,Jang2011,McCutcheonNazir2010,mccutcheon11,KolliNazir2011,mccutcheon11b} From this we may extract the relevant ET rates, extending previous analyses to a broader set of parameters and looking in more detail at the assumptions required for the MJ rates to be valid; in fact, these rates arise naturally from our master equation in the semiclassical 
limit where the temperature is high 
compared to the energy scales of the environment. In general, we find that the dynamics of donor-acceptor (DA) populations predicted by our master equation can differ considerably from that given by the simpler MJ rates. The main reason for this discrepancy 
is the assumption inherent to the MJ analysis of exponential ET from donor to acceptor, which we find to be invalid 
for a wide set of parameter values. 

Our approach also has the crucial advantage of allowing us to 
additionally incorporate 
the effects of 
dissipation on the odorant mode, since it is treated explicitly in our formalism. We shall show that by introducing sufficiently strong mode dissipation, 
it is possible to bring the DA dynamics derived from our  master equation into a simpler exponential form. 
However, even within this strongly dissipative regime, our master equation predictions can still differ markedly 
from the semiclassical MJ theory, depending upon the specific nature of the environment experienced by the DA pair. In particular, we show that while in the MJ case 
the receptor is very sensitive to the presence or absence of the odorant, it is far less so to the specific odorant vibrational frequency. We find, however, that the frequency resolution can be enhanced by considering environments which contain components of similar or larger energy to that set by the ambient temperature, i.e.~by working outside the semiclassical 
limit. 
More generally, we show that when considering odorant mode dissipation, both the 
sensitivity of the switch to the presence or absence of the odorant, and to the resonance conditions between the odorant and the DA pair, can be significantly amplified 
in comparison to the dissipationless case. We thus find that odorant dissipation plays a constructive role in 
enhancing the vibrational sensing capabilities 
of our molecular switch. 

This paper is structured as follows. We begin in Section~\ref{sec:model} by describing the molecular switch model and giving an estimate of the parameters relevant to our analysis. In Section~\ref{sec:incoherent} we briefly outline the derivation of our polaron-representation master equation, with further details given in the Appendix. Sections~\ref{sec:et} A and B are devoted to showing how the MJ rates naturally emerge in the semiclassical limit of our master equation, in the absence and presence of the odorant, respectively. Section~\ref{sec:et} C contrasts the full master equation dynamics with a simpler rate analysis in the illustrative case of a dissipationless odorant mode, while Section~\ref{sec:nonperturb} considers the impact of odorant dissipation upon the ET dynamics and the resulting switch frequency resolution. Finally, we summarise our findings in Section~\ref{sec:conc}. 

\section{Model of olfaction}
\label{sec:model}

\begin{figure}[t]
\includegraphics[width=0.38\textwidth]{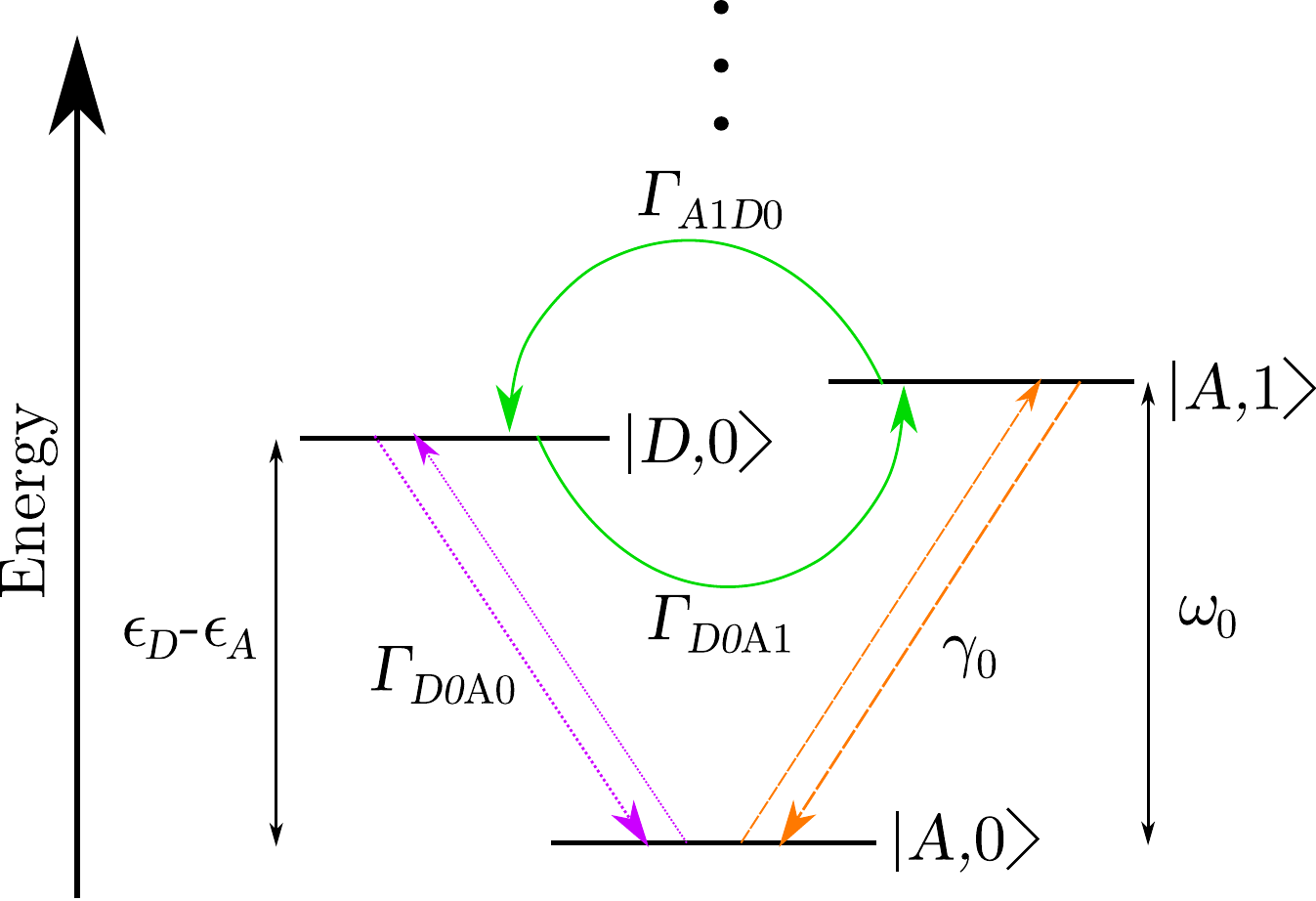}
\caption{{\bf Relevant DA-odorant energy levels.} The three most relevant single-electron joint DA-odorant states and the principal transition rates between them. The rates  $\Gamma_{DnAm}$ and $\Gamma_{AnDm}$ are derived in Sec.~\ref{sec:et}, whilst odorant dissipation with rate $\gamma_0$ is discussed in Sec.~\ref{sec:nonperturb}. The notation $\ket{X,n}$ refers to the joint state of the DA pair, $X\in\{D,A\}$, and the odorant in Fock state $\ket{n}$.}
\label{energylevels}
\end{figure}

Few details are presently known about olfactory receptors and their properties.~\cite{buck_novel_1991,axel_scents_2005,buck_unraveling_2005} Here, following earlier work,~\cite{brookes_could_2007,solovyov_vibrationally_2012,bittner_quantum_2012} we study a simplified model which captures the essential physics of the electron transfer process. We assume that there exist specific electronic states of the receptor that can be identified as a DA pair, with other levels well separated in energy. 
We model the vibrationally-assisted ET process 
using a spin-boson Hamiltonian. The DA pair is coupled to an environment represented by a bath of harmonic oscillators (which includes the vibrational degrees of freedom of the receptor) and to the odorant (when present in the receptor). The odorant is modelled as a single mode harmonic oscillator, though this could also be extended to a set of modes. 
Our Hamiltonian thus takes the form
\begin{align}
H&=\epsilon_D\ket{D}\bra{D}+\epsilon_A\ket{A}\bra{A}+V\big(\ket{A}\bra{D}+\ket{D}\bra{A}\big)\nonumber\\
&+\omega_0a^\dg a+(\gamma_D\ket{D}\bra{D}+\gamma_A\ket{A}\bra{A})(a^\dg+a)\nonumber\\
&+\sum_k\big[\omega_kb^\dg_k b_k+(\gamma_{kD}\ket{D}\bra{D}+\gamma_{kA}\ket{A}\bra{A})(b^\dg_k+b_k)\big].\label{initial_h}
\end{align}
Here, $\ket{X}$, with $X=D,A$, represents the donor ($D$) and acceptor ($A$) electronic state with on-site energy $\epsilon_X$, and $V$ refers to the tunnel coupling. The odorant molecular mode, with creation (annihilation) operator $a^{\dagger}$ ($a$), has frequency $\omega_0$ and is coupled to the DA pair via $\gamma_X$. The environmental oscillators, with creation (annihilation) operators $b_k^\dg$ ($b_k$) for modes of frequency $\omega_k$, couple to the receptor 
electronic sites via $\gamma_{kX}$. We shall consider odorant dissipation in Section~\ref{sec:nonperturb} below. Transfer of an electron gives rise to changes in the local electric field and justifies the present form of interaction between the oscillators and the DA pair. \cite{brookes_could_2007,solovyov_vibrationally_2012,bittner_quantum_2012,nitzan_chemical_2006} Fig.~\ref{energylevels} shows the most relevant energy levels of the joint system along with transition rates between them (to be derived in later sections). Specific parameter values for the model
will be discussed below. 

In Refs.~\onlinecite{brookes_could_2007,solovyov_vibrationally_2012}, MJ formulas were employed to define two types of ET rate: elastic, in the absence of the odorant, and inelastic, in which the odorant is present. These rates were calculated using Fermi's golden rule, wherein the tunnelling term $V$ is treated as a perturbation. This naturally gives rise to a representation of the receptor-odorant-environment basis in terms of displaced oscillator states---as can be seen from Eq.~(\ref{initial_h}) by setting $V=0$---between which it is assumed that incoherent ET occurs. This is, in fact, a reflection of the polaron picture that we shall discuss in the next section. In Ref.~\onlinecite{brookes_could_2007}, typical ET timescales were estimated to be $\sim100$~ns (elastic) versus $\sim1$~ns (inelastic), predicting the inelastic process to be much faster than the elastic one, as required for discriminating between the presence and the absence of the odorant. Ref.~\onlinecite{solovyov_vibrationally_2012} also discusses a wider range of parameters in which a separation of timescales between the elastic and inelastic processes can be obtained.  

Here, rather than following a Fermi golden rule calculation, 
we shall look instead at receptor ET dynamics and issues such as odorant vibrational frequency resolution from the perspective of the master equation formalism. This includes the MJ rate analysis as a limiting case, but can also go well beyond the restricted regime of validity of such an approach. 

\subsection{Estimated parameters}

For consistency with the published literature, we follow the estimation of parameters presented in Refs.~\onlinecite{brookes_could_2007,solovyov_vibrationally_2012}. 
In common with previous studies, we assume that the energy gap $\epsilon_D-\epsilon_A$ is relatively close to resonance with the odorant vibrational frequency $\omega_0$. 
We shall, however, explore an effective window of frequencies away from resonance to study the issue of 
molecular frequency recognition, i.e.~we would like to explore whether the switch is sensitive merely to the presence or absence of the odorant, or to its particular vibrational frequency as well. Typical values for $\omega_0$ are in the range of $70-400$~meV, and we choose $\epsilon_D-\epsilon_A = 200$~meV for the DA energy gap. The DA tunnel coupling is estimated to be on the order of $V\sim1$~meV.~\cite{brookes_could_2007} Increasing $V$ acts to enhance the ET rate in the absence of the odorant in the receptor, which is disadvantageous as far as the switching mechanism 
is concerned. Keeping $V$ small in comparison to other system parameters is therefore well motivated physically. Estimates of the coupling between the DA pair and the odorant mode, and of the reorganization energy of the multi-mode environment, 
have been given as $(\gamma_D-\gamma_A)^2/\omega_0^2\sim0.01$ and $\lambda\sim30$~meV, respectively. Here, we shall explore variations in the DA-environment coupling strength through the reorganisation energy, focussing on the range $\lambda\sim10-60$~meV. 
Previously, the detailed properties of the environment, such as the spectral density, 
\begin{equation}\label{spectraldensity}
J(\omega)=\sum_k(\gamma_{kD}-\gamma_{kA})^2\delta(\omega-\omega_k),
\end{equation}
were not discussed, as they do not enter the semiclassical MJ rates. They are, however, important for our more general 
analysis. We choose to work with an Ohmic environment, $J(\omega)\propto \omega$ as $\omega \rightarrow 0$, with a characteristic high frequency cut-off $\omega_c$, since in the absence of more detailed information, it straightforwardly reproduces the results of Refs.~\onlinecite{brookes_could_2007,solovyov_vibrationally_2012} in the semiclassical 
limit ($\beta\omega_c\ll1$, where $\beta=1/k_BT$ is the inverse temperature). We take the cut-off to be exponential, leading to the following spectral density defined also in terms of the reorganisation energy:
\begin{equation}
J(\omega)= \lambda \frac{\omega}{\omega_c} {\rm e}^{-\omega/\omega_c}.\label{eq:expspecden}
\end{equation} 
We shall vary the ratio $\omega_c / k_BT$ 
from low to high 
in order to explore the effect of widening the range of frequencies within the environment that can interact with the receptor DA pair. We assume that $T=300$~K throughout, 
and summarise the model parameters in Table~\ref{partable}.
\begin{table}
\begin{center}
\begin{tabular}[t]{| c | c | c | c |}
\hline
Parameter & Value & Parameter &Value \\
\hline
$\epsilon_D-\epsilon_A$ & $200$~meV & $V$ & $1$~meV\\
\hline
$\omega_0$ & $100-300$~meV 
& $\lambda$ & $10-60$~meV\\
\hline
$(\gamma_D-\gamma_A)^2/\omega_0^2$ & $0.01$ & $\omega_c$ & $\frac{k_BT}{10}$, $k_BT$, $2k_BT$\\ 
\hline
$T$ & $300$~K & $k_BT$ & $25.85$~meV\\ 
\hline
\end{tabular}\caption{Table of parameter values used in this article.\label{partable}}
\end{center}
\end{table}

\section{Polaron master equation} 
\label{sec:incoherent}

As noted, for our system to act as an effective molecular switch, the Hamiltonian parameters should be such that unwanted transitions from donor to acceptor are avoided when the odorant is absent. This requires the tunnel coupling $V$ to be small compared to other energy scales in the problem. In this regime, it is convenient to move into a polaron transformed reference frame to remove the linear coupling terms in Eq.~(\ref{initial_h}). Provided $V$ is indeed small, then perturbative expansions in the transformed basis are valid over a much wider range of system-environment coupling strengths than those in the untransformed case.~\cite{holstein_studies_1959-1,holstein_studies_1959,JacksonSilbey1983,jang_theory_2008,nazir09,jang09,Jang2011,McCutcheonNazir2010,mccutcheon11,KolliNazir2011,mccutcheon11b} 
Let us consider the (unitary) polaron transformation $U = e^{- (S_\mathcal{A}+S_\mathcal{B})}$ acting on our Hamiltonian, where
\begin{align}
S_\mathcal{A}=\big(u_D\ket{D}\bra{D}+u_A\ket{A}\bra{A}\big)(a^\dg-a),
\end{align}
and
\begin{align}
S_\mathcal{B}=\sum_k\big(\alpha_{kD}\ket{D}\bra{D}+\alpha_{kA}\ket{A}\bra{A}\big)(b_k^\dg-b_k),
\end{align}
with $u_X=\gamma_X/\omega_0$ and $\alpha_{kX}=\gamma_{kX}/\omega_k$. 
The polaron transformed Hamiltonian takes the form $H_{\rm P}=U^{\dagger}HU=H_{\rm SP}+H_B+H_{\rm IP}$, where
\begin{align}\label{HSP}
H_{{\rm SP}}&=\epsilon_D'\ket{D}\bra{D}+\epsilon_A'\ket{A}\bra{A}+\omega_0a^\dg a,
\end{align}
\begin{align}
H_B=\sum_k\omega_kb_k^\dg b_k,
\end{align}
and
\begin{align}\label{HIP}
H_{\rm IP}&=V\big(\ket{D}\bra{A}\mathcal{A}_+\mathcal{B}_++\ket{A}\bra{D}\mathcal{A}_-\mathcal{B}_-\big).
\end{align}
Here, $\epsilon_X'=\epsilon_X-\omega_0u_X^2-\sum_k\omega_k\alpha_{kX}^2$ is the polaron-shifted on-site energy, and $\mathcal{A}_{\pm}=\mathcal{D}(\pm(u_D-u_A))$ and $\mathcal{B}_\pm=\Pi_k\mathcal{D}_k(\pm(\alpha_{kD}-\alpha_{kA}))$ are the new oscillator interaction operators, written in terms of the displacement operators $\mathcal{D}(u)=e^{ua^\dg-u^*a}$ and $\mathcal{D}_k( \alpha_k)=e^{\alpha_kb_k^\dg-\alpha_k^*b_k}$, respectively. Note that the polaron transformation leaves the operators $\ket{D}\bra{D}$ and $\ket{A}\bra{A}$, required for calculating DA populations, unchanged. 

Tracing out the environment and treating the perturbative term $H_{\rm IP}$ up to second order in the standard Born-Markov approximations,~\cite{breuer_theory_2002} we obtain a 
master equation in the polaron frame interaction picture (with respect to $H_{\rm SP}+H_B$), given by
\begin{align}
\dot{\tilde{\rho}}_{\rm SP}(t)&=-V^2\int_0^\infty d\tau\nonumber\\
&\big\{\big([\ket{A}\bra{D}\mathcal{A}_-(t),\ket{D}\bra{A}\mathcal{A}_+(t-\tau)\tilde{\rho}_{\rm SP}(t)]e^{-i\epsilon \tau}\nonumber\\
&+[\ket{D}\bra{A}\mathcal{A}_+(t),\ket{A}\bra{D}\mathcal{A}_-(t-\tau)\tilde{\rho}_{\rm SP}(t)]e^{i\epsilon\tau}\big)C(\tau)\nonumber\\
&+\mathrm{H.c.}\big\},\label{me}
\end{align}
see 
the Appendix 
for further details. 
Here, $\tilde{\rho}_{\rm SP}(t) = {\rm tr}_B\{\tilde{\chi}_{\rm P}(t)\}$ is the reduced density operator describing the DA pair and odorant mode on tracing out the receptor environment, with $\tilde{\chi}_{\rm P}(t)$ the total density operator in the polaron frame interaction picture. The bath correlation function, defined as $C(\tau)={\rm tr}_{B}\left(\mathcal{B}_{\pm}(\tau)\mathcal{B}_{\mp}(0)\rho_B\right)$ with $\mathcal{B}_{\pm}(t)=e^{iH_Bt}\mathcal{B}e^{-iH_Bt}$, takes the form of an exponential of the lineshape function:
\begin{equation}
C(\tau)=e^{-\int_0^\infty d\omega\frac{J(\omega)}{\omega^2}[(1-\cos\omega\tau)\coth (\beta\omega/2)+i\sin \omega\tau]}, \label{eq:corrfuncs}
\end{equation}
with $\epsilon=\epsilon_D'-\epsilon_A'$ now the DA energy gap. 

\section{Electron Transfer dynamics} 
\label{sec:et}

Having derived our polaron master equation, we shall use it 
below, in Section C, to investigate the receptor ET dynamics 
beyond the MJ rate analysis employed in earlier work. First, however, we show how the MJ rates arise 
from our master equation in the semiclassical limit, for both 
elastic and inelastic processes, and thus how our theory is consistent with previous treatments in this regime. In addition, we discuss other important transition rates which we shall subsequently show are suppressed only with the introduction of strong dissipation acting on the odorant mode.

\subsection{Odorant absent} 

Let us consider the population transfer from donor to acceptor for the case in which the odorant is absent, 
meaning that $\mathcal{A}_\pm(t)=\openone$ in Eq.~(\ref{me}). We can then easily derive rate equations governing the donor ($p_D(t)$) and acceptor ($p_A(t)$) population dynamics. Our interest lies in the rates that appear in these equations
\begin{equation}
\Gamma(\pm\epsilon)\,=\,\int_{-\infty}^{\infty}d\tau e^{\pm i\epsilon\tau}C(\tau),
\end{equation}
where, by defining $\Gamma_{DA}=V^2\Gamma(\epsilon)$ and $\Gamma_{AD}=V^2\Gamma(-\epsilon)$, we obtain
\begin{align}
\dot{p}_{D}(t)&=-\Gamma_{DA}p_{D}(t)\,+\,\Gamma_{AD}p_{A}(t),\nonumber\\
\dot{p}_{A}(t)&=-\Gamma_{AD}p_{A}(t)\,+\,\Gamma_{DA}p_{D}(t).\nonumber
\end{align}
In the limit that $\Gamma_{DA}\gg\Gamma_{AD}$, these equations define exponential transfer of population from donor to acceptor at (approximately) the rate $\Gamma_{DA}$, 
which we write as 
\begin{align}\label{dageneralodorantabsent}
\Gamma_{DA}=V^2\int_{-\infty}^{\infty}d\tau e^{i\epsilon\tau}e^{-\varphi(\tau)},
\end{align}
where
\begin{align}\label{varphi}
\varphi(\tau)=\int_0^\infty d\omega\frac{J(\omega)}{\omega^2}\big[(1-\cos\omega\tau)\coth(\beta\omega/2)+i\sin\omega\tau\big].
\end{align}
For a low-frequency environment in which $\beta\omega_c\ll1$, we may derive a simple form for $\Gamma_{DA}$ which turns out to be the same as the MJ rate.  
Taking the spectral density defined in Eq.~(\ref{eq:expspecden}) 
and expanding $\coth(\beta\omega/2)\approx2/\beta\omega$ in Eq.~(\ref{varphi}), we find
\begin{align}
\varphi(\tau)&\approx\frac{\lambda}{\beta\omega_c^2}\left[\omega_c(i\beta+2\tau)\tan^{-1}(\omega_c\tau)-\ln(1+\omega_c^2\tau^2)\right].
\end{align}
In the regime in which we are presently interested, $e^{-\varphi(\tau)}$ is strongly peaked around $\tau=0$, such that we may expand $\varphi(\tau)$ to second order in $\tau$ to give
\begin{align}
\varphi(\tau)&\approx\,i\lambda\tau\,+\,\frac{\lambda\tau^2}{\beta}.
\end{align}
With these assumptions we can write
\begin{align}\label{daodorantabsent}
\Gamma_{DA}&\approx\,V^2\int_{-\infty}^\infty d\tau e^{i(\epsilon-\lambda)\tau}e^{-\lambda\tau^2/\beta}\nonumber\\
&=2\pi V^2\frac{1}{\sqrt{4\pi k_BT\lambda}}\exp\bigg[-\frac{(\epsilon-\lambda)^2}{4k_BT\lambda} \bigg], 
\end{align}
which agrees with the elastic rate (odorant absent) presented in Ref.~\onlinecite{brookes_could_2007} when $\alpha_{kD}=-\alpha_{kA}$ is assumed, and as a consequence $\epsilon\rightarrow\epsilon_D-\epsilon_A$. We apply this constraint on $\alpha_{kX}$ throughout the paper, although little modification would be required to account for the more general case. Choosing $\lambda=30$~meV, and other parameters as outlined in Table~\ref{partable}, 
we obtain $\Gamma_{DA}=5.67\times10^{-6}$~meV from Eq.~(\ref{daodorantabsent}), which corresponds to $\tau_{DA}=1/\Gamma_{DA}=116$~ns as the DA transfer time in the absence of the odorant. 

Of course, the limit $\beta\omega_c\ll1$ used to derive Eq.~(\ref{daodorantabsent}) may not always be met, in which case we can use the more general form of Eq.~(\ref{dageneralodorantabsent}) to define the DA elastic transfer rate. 
Importantly, this allows us to discuss lower temperatures and larger environmental cut-offs than those to which the MJ rates apply.

\subsection{Odorant present} 

A similar rate can also be derived when the odorant is present in the receptor. 
In this situation, we must deal with a more complex system, as the reduced density operator (after tracing out the environmental degrees of freedom) now encompasses both the two-level DA pair and the odorant harmonic oscillator. Obtaining general equations of motion for the DA populations 
becomes significantly more involved. However, this is unnecessary if we instead look at ET rates between specific states of the combined DA-odorant system. 
Let us assume that we initialize the system in the state $|D,0\rangle$ (electron on the donor, odorant in its ground-state $|0\rangle$) and we are interested in the rate of ET to a state of the form $|A,n\rangle$, where $|n\rangle$ is an arbitrary Fock (number) state of the odorant. We are thus considering situations in which the receptor population transfer $|D\rangle\rightarrow|A\rangle$ is accompanied by 
excitation of the odorant vibrational mode $|0\rangle\rightarrow|n\rangle$, 
which may or may not act to enhance the rate associated with the process.

We return to the master equation [Eq.~(\ref{me})] to derive an expression for the dynamics of the population of the acceptor and a given Fock state $|n\rangle$ of the odorant, $\tilde{\rho}_{AnAn}={\rm tr}_{S+O}(|A,n\rangle\langle n,A|\tilde{\rho}_{\rm SP}(t))=\langle n, A|\tilde{\rho}_{\rm SP}(t)|A,n\rangle$, where the trace is taken over both the odorant ($O$) and the DA pair ($S$) degrees of freedom. Decomposing the DA-odorant density operator as $\tilde{\rho}_{\rm SP}(t)=\sum_{X,X',l,m}\tilde{\rho}_{XlX'm}(t)|X,l\rangle\langle X',m|$, where $X,X'\in\{D,A\}$ and $l,m$ are odorant Fock states, allows us to identify the different contributions to the change in population of the state $|A,n\rangle$. In particular, if we assume that the only donor-odorant state of interest is one with no odorant excitations, $|D,0\rangle$ --- valid when we initialise the system in this state and energy splittings are large enough to suppress transitions to any other donor-odorant states --- we may then define a rate of transfer $|D,0\rangle\rightarrow|A,n\rangle$ as 
  \begin{align}\label{polrateodorant}
  \Gamma_{D0An}=&\int_0^\infty d\tau\bigg(e^{i\epsilon\tau}C(\tau)\langle n|\mathcal{A}_-(t-\tau)|0\rangle\langle0|\mathcal{A}_+(t)|n\rangle\nonumber\\
  &+e^{-i\epsilon\tau}C^*(\tau)\langle n|\mathcal{A}_-(t)|0\rangle\langle 0|\mathcal{A}_+(t-\tau)|n\rangle\bigg){V^2}.
  \end{align}
Using $\langle n|D(\alpha)|0\rangle=(\alpha^n/\sqrt{n!})e^{-|\alpha|^2/2}$, it is straightforward to calculate the expectation values of $\mathcal{A}_\pm(t)$. 
We then arrive at 
 \begin{align}
 \Gamma_{D0An}=&\frac{V^2(u_D-u_A)^{2n}e^{-|u_D-u_A|^2}}{n!}\nonumber\\
 &\times\int_{-\infty}^\infty d\tau e^{i(\epsilon-n\omega_0)\tau}e^{-\varphi(\tau)}, \label{eq:Ahsansrate}
 \end{align}
which generalises Eq.~(\ref{dageneralodorantabsent}) in the presence of the odorant ($\Gamma_{DA}=\Gamma_{D0A0}$).
At this point we can again take the limit $\beta\omega_c\ll1$, and follow the same steps that led us from Eq.~(\ref{dageneralodorantabsent}) to Eq.~(\ref{daodorantabsent}) to find
\begin{align}\label{daodorantpresent}
\Gamma_{D0An}\approx&\;2\pi V^2\frac{(u_D-u_A)^{2n}e^{-|u_D-u_A|^2}}{n!\sqrt{4\pi k_BT\lambda}}\nonumber\\
&\times\exp\bigg[-\frac{(\epsilon-n\omega_0-\lambda)^2}{4k_BT\lambda}\bigg].
\end{align}
Once again, this result for the inelastic ET rates ($n>0$) agrees with the MJ form 
found in Ref.~\onlinecite{brookes_could_2007} 
when we assume $u_D=-u_A$ and $\alpha_{kD}=-\alpha_{kA}$, such that $\epsilon\rightarrow\epsilon_D-\epsilon_A$. Using the values in Table~\ref{partable}, choosing $\lambda=30$~meV and $\omega_0=\epsilon_D-\epsilon_A$, we obtain $\Gamma_{D0A1}=4.71\times10^{-4}$~meV, which corresponds to a transfer time of $\tau_{D0A1}=1/\Gamma_{D0A1}=1.4$~ns. This is far shorter than the $116$~ns transfer time found above in the absence of the odorant, as required for a viable molecular switch. Additionally, if we take the same parameter values and look at the rate for the two-phonon transition we then find $\Gamma_{D0A2}=1.24\times10^{-13}$~meV, and a corresponding time $\tau_{D0A2}=5.3$~s, confirming that the single-phonon process is dominant within this treatment. As before, in situations in which the limit $\beta\omega_c\ll1$ does not apply, 
we may instead use Eq.~(\ref{eq:Ahsansrate}) to define the inelastic rates, again generalising the MJ rates to a wider range of parameters.

Of course, our master equation also allows us to calculate the reverse rates $\Gamma_{AnD0}$ (as well as other rates such as $\Gamma_{DnDm}$ and $\Gamma_{AnAm}$, which do not play a major role). Following a derivation similar to that leading to Eq.~(\ref{eq:Ahsansrate}), we find
\begin{align}
 \Gamma_{AnD0}=&\frac{V^2(u_D-u_A)^{2n}e^{-|u_D-u_A|^2}}{n!}\nonumber\\
 &\times\int_{-\infty}^\infty d\tau e^{i(\epsilon-n\omega_0)\tau}e^{-\varphi(-\tau)}, \label{eq:backrate}
\end{align}
and, in the 
limit $\beta\omega_c\ll1$, 
\begin{align}\label{adodorantpresent}
\Gamma_{AnD0}\approx&\;2\pi V^2\frac{(u_D-u_A)^{2n}e^{-|u_D-u_A|^2}}{n!\sqrt{4\pi k_BT\lambda}}\nonumber\\
&\times\exp\bigg[-\frac{(\epsilon-n\omega_0+\lambda)^2}{4k_BT\lambda}\bigg].
\end{align}
Notably, the single-phonon reverse transfer rate, $\Gamma_{A1D0}$, is equal to the donor-to-acceptor rate, $\Gamma_{D0A1}$, when the odorant is resonant with the receptor ($\omega_0 = \epsilon_D-\epsilon_A$). We shall see below that this has important implications for the dynamics of the DA pair over a wide range of parameters, often invalidating the treatment of inelastic ET as being a one-way process. 

\subsection{Master equation dynamics versus ET rates}

\begin{figure*}[t]
\includegraphics[width=0.73\textwidth]{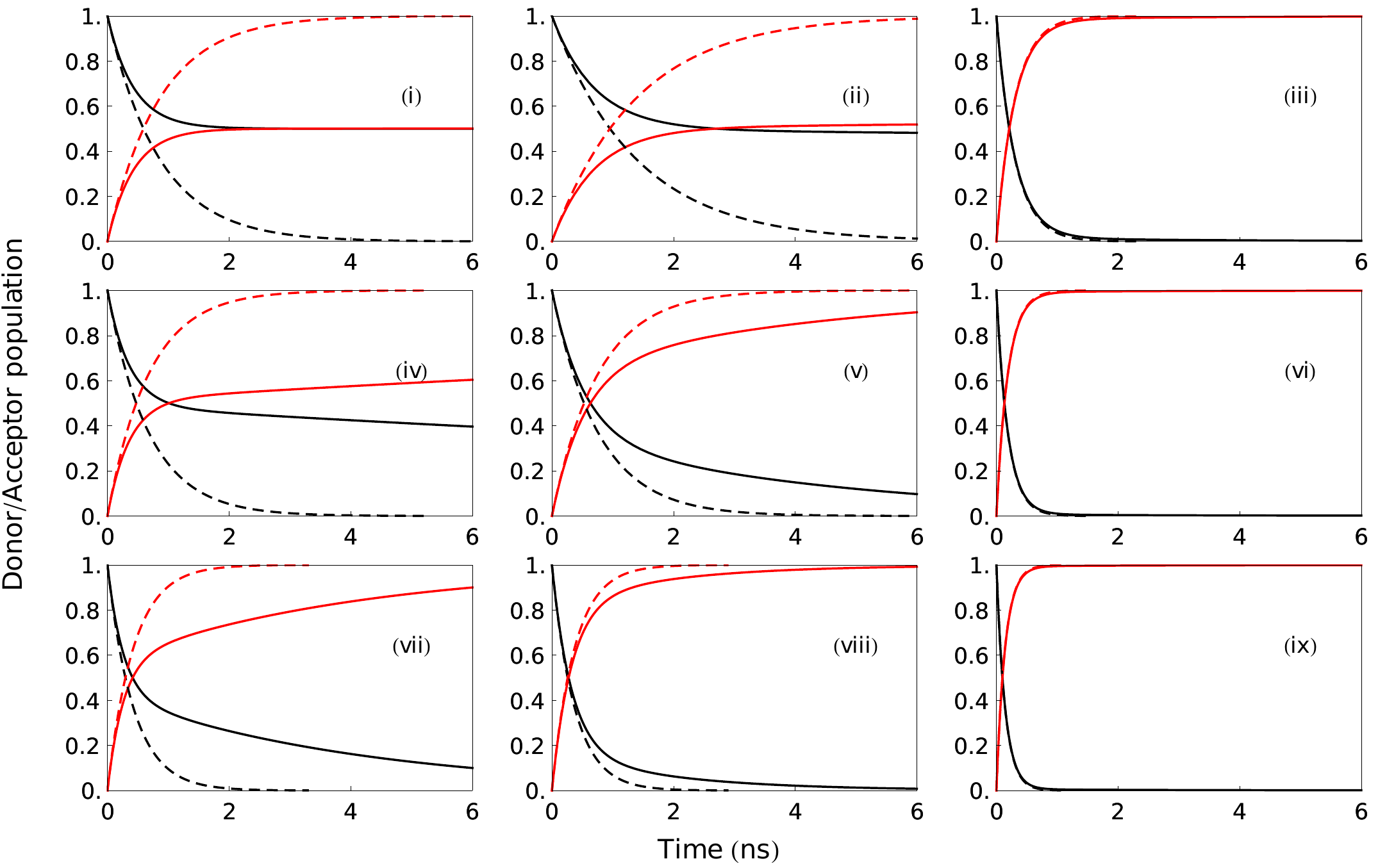}
\caption{{\bf Donor and acceptor population dynamics.} Comparison of the donor, $p_D(t)$ (black, $p_D(0)=1$), and acceptor, $p_A(t)$ (red, $p_A(0)=0$), population dynamics as predicted by the polaron master equation~(\ref{me}) (solid lines, calculated using $3$ odorant states to ensure numerical convergence) and by the total donor-to-acceptor  
ET rate $\sum_{n=0}^3 \Gamma_{D0An}$ from Eq.~(\ref{eq:Ahsansrate}) (dashed lines). 
The reorganisation energy $\lambda$ increases from left to right: $\lambda = 15$~meV in the left column, $\lambda = 30$~meV in the middle column and $\lambda = 60$~meV in the right column. The bath cut-off frequency $\omega_c$ increases from top to bottom, with $\omega_c=k_BT/10$ for panels (i - iii), $\omega_c=k_BT$ for panels (iv - vi) and $\omega_c=2k_BT$ for panels (vii - ix). The odorant frequency is chosen to be resonant with the DA pair, i.e. $\omega_0=\epsilon_D-\epsilon_A=200$~meV, 
and other parameters are as listed in Table~\ref{partable}. Note that for each set of parameters, the populations eventually tend to a thermal distribution with respect to $H_{DA} = {\epsilon '}_D\ket{D}\bra{D}+ {\epsilon '}_A\ket{A}\bra{A}$ in the steady state; this is true even for panels (i) and (ii), which equilibrate on a very long time scale.} \label{rates-dynamics}
\end{figure*} 

We are now in a position to compare our full polaron master equation [Eq.~(\ref{me})] with a model in which the donor population simply decays with the ET rates calculated in the previous section. Specifically, we would like to know how reliably we can apply a rate analysis, such as the MJ evaluation employed  
in previous studies,~\cite{brookes_could_2007,solovyov_vibrationally_2012}  
to parameter regimes estimated to be relevant to the olfactory process.
We have already seen that with the odorant present, the possibility of transfer from acceptor to donor 
cannot be neglected when close to resonance. This is especially true when the inelastic 
process dominates, 
as the reverse rate in the elastic case is suppressed by a factor $\sim\exp{[-(\epsilon_D-\epsilon_A)\lambda/k_BT]}$. 
Without further assumptions, we thus expect exponential decay from donor to acceptor to hold 
only when the ET 
is mediated primarily by the environment ($n=0$), 
rather than by the odorant itself. 

This intuition is borne out in Fig.~\ref{rates-dynamics}, which shows a comparison between the DA population dynamics (in the presence of the odorant) 
calculated from our master equation~(\ref{me}), and 
predicted by the ET rates of 
Eq.~(\ref{eq:Ahsansrate}). The receptor is taken to be initialised in the donor state, with 
the odorant mode and the environment in thermal states, of $H_{\rm O}=\omega_0a^{\dagger}a$ and of $H_B$, respectively.~\cite{Note2}
%\footnote{For $\omega_0=200$~meV and $T=300$~K this corresponds to the odorant initial state being essentially the ground state, and is therefore consistent with the initial condition in the rate analysis.}. 
As can clearly be seen, the ET dynamics 
can differ considerably between 
the two methods. As expected, the best agreement is found for large reorganisation energies (right column), where elastic transfer due to the environment 
dominates over inelastic transfer via the odorant. 
On the other hand, when the odorant does play a significant role in mediating the ET, then the agreement is generally poor, with our master equation predicting DA dynamics that cannot be fitted by a single exponential form.~\cite{Note3} 
%\footnote{The DA dynamics is generally well fitted by a biexponential form.}. 
This is particular true in the low-frequency environment limit ($\omega_c \ll k_B T$) shown in panels (i) and (ii), to which the MJ rates would usually be assumed to apply.~\cite{Note4}
%\footnote{Note that the ET rates used to calculate the dashed curves in panels (i - iii) of Fig.~\ref{rates-dynamics} do agree with the MJ rates.}. 
These discrepancies 
suggest, in fact, that it is the combination of population 
accumulation within the odorant and the inherent competition between forward and backward processes that 
limits the timescale for complete transfer from donor to acceptor in the master equation. 

\begin{figure*}[t]
\includegraphics[width=0.73\textwidth]{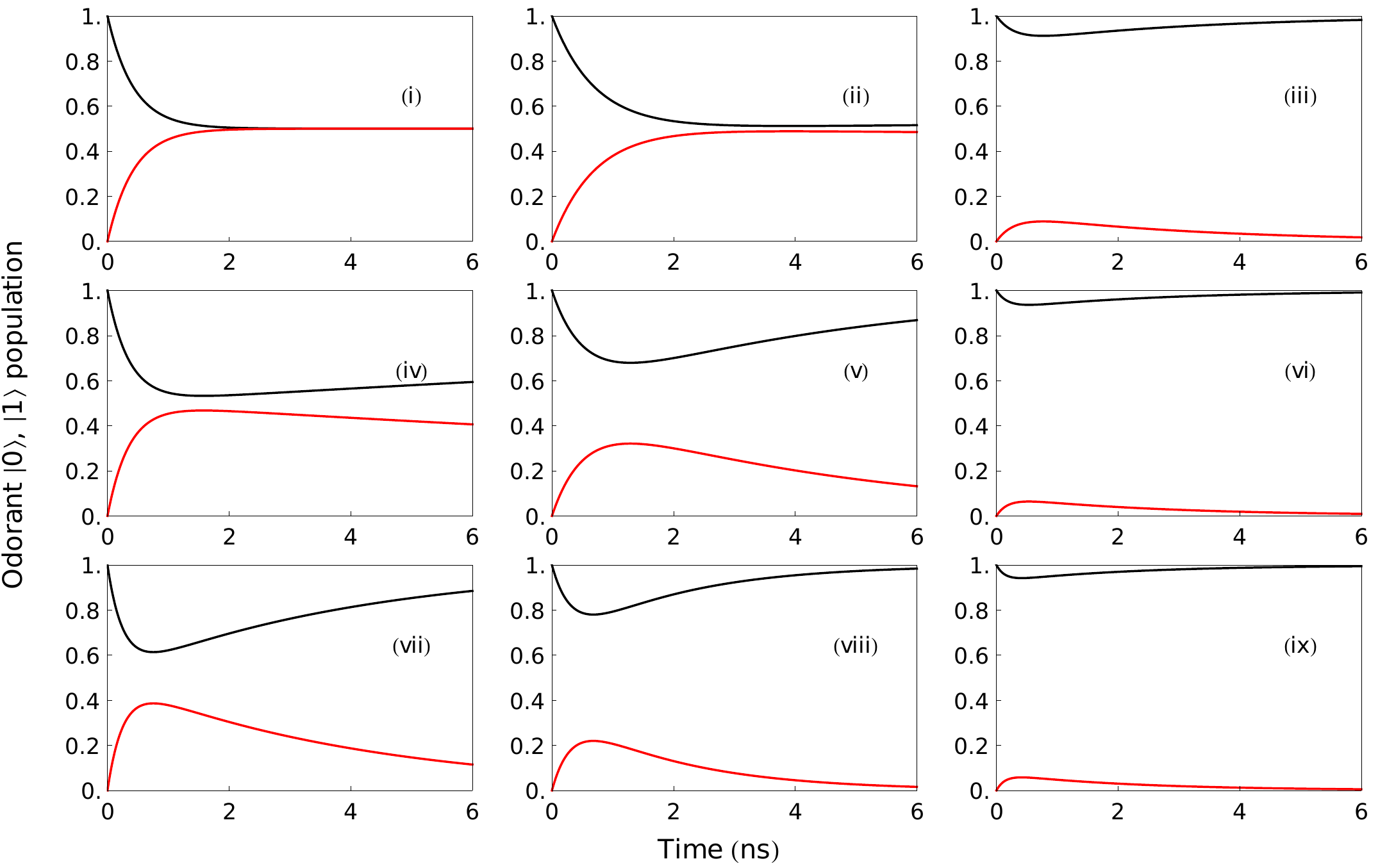}
\caption{{\bf Odorant dynamics.} Populations of the odorant ground state, $p_{\ket{0}\bra{0}}(t)$ (black, $p_{\ket{0}\bra{0}}(0)=1$), and first excited state, $p_{\ket{1}\bra{1}}(t)$ (red, $p_{\ket{1}\bra{1}}(0)=0$), calculated from Eq.~(\ref{me}). The reorganisation energy $\lambda$ increases from left to right: $\lambda = 15~\rm meV$ in the left column, $\lambda =~30 \rm meV$ in the middle column and $\lambda = 60~\rm meV$ in the right column. The bath cut-off frequency $\omega_c$ increases from top to bottom, with $\omega_c=k_BT/10$ for panels (i - iii), $\omega_c=k_BT$ for panels (iv - vi) and $\omega_c=2k_BT$ for panels (vii - ix). The odorant frequency is $\omega_0=\epsilon_D-\epsilon_A=200$~meV. Populations of higher Fock states are negligible at all times.}
\label{osc-dyn}
\end{figure*} 

We can thus gain further insight by looking at the population dynamics of the odorant mode. 
In Fig.~\ref{osc-dyn} we plot the evolution of the lowest two odorant states as predicted by our master equation, for the same parameter sets as in Fig.~\ref{rates-dynamics}. Note that other levels are never significantly populated, since the total initial energy is insufficient to further excite the mode. In the cases in which the odorant-assisted (inelastic) rate dominates, excitation is quickly transferred from the DA pair to the odorant. 
The electron is then equally likely to be found on the acceptor (having deposited energy to the odorant mode) as it is on the donor. Due to the equality of the forward and backward rates on resonance, the only way energy can now exit the system, to allow full ET 
to the acceptor, is into the environment, the timescale for which is dependent upon the reorganisation energy and cut-off frequency. Since the rate 
analysis that leads to the dashed curves in Fig.~\ref{rates-dynamics} ignores the possibility of reverse acceptor-to-donor transfer, it consequently overestimates the inelastic transfer rate. It is worth noting that even for the cases in which the master equation dynamics shown in Fig.~\ref{rates-dynamics} 
%~and~\ref{osc-dyn} 
appears to reach a (quasi) steady state with the donor and acceptor similarly 
populated, in fact full 
ET from donor to acceptor does still eventually occur, albeit on a very long timescale.

Of course, in any realistic physical setting, the odorant mode would not be isolated from the environment. In MJ theory, for the semiclassical limit to be valid in the presence of the odorant an implicit assumption is made that the odorant vibrational mode 
dissipates its energy to the environment on a timescale much shorter than that of the ET process. However, within our master equation approach we can 
explicitly include mode dissipation, and indeed explore the effects of varying the associated rate. 
Thus, by contrasting the dynamics illustrated here in the 
absence of odorant dissipation 
with that in its presence 
in the next section, we can investigate the detailed role such dissipation plays in vibrationally-assisted ET. We shall  
show that for strong dissipation we can recover exponential 
inelastic ET, though generally at a rate that differs from the semiclassical MJ formula. 
Furthermore, we shall demonstrate that 
reducing the level of odorant dissipation can in fact act to limit the frequency detection properties of the receptor, suggesting that strong dissipation may actually be beneficial for obtaining frequency selective switching processes. 

\section{Dissipation assisted electron transfer}
\label{sec:nonperturb}

\begin{figure}
\centering
\includegraphics[width=0.45\textwidth]{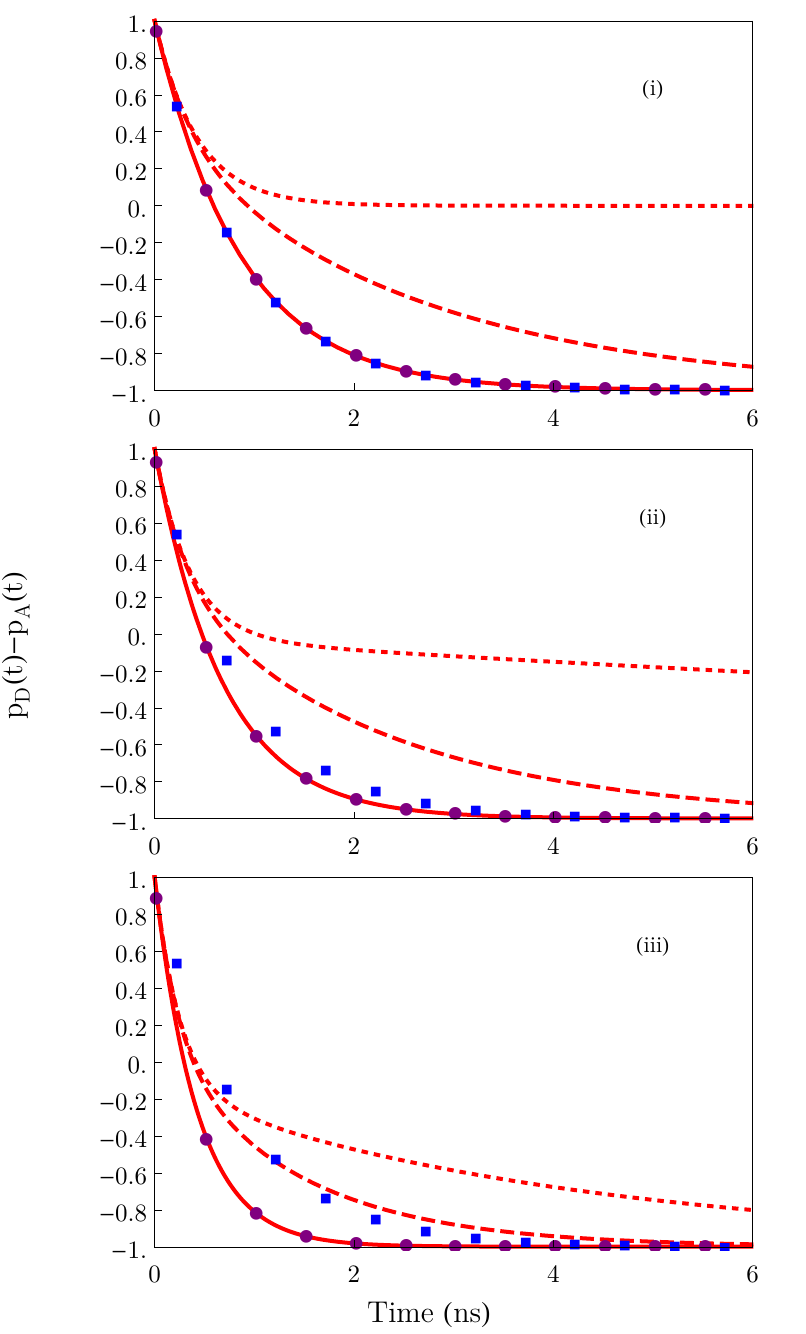}
\caption{{\bf Effect of odorant mode dissipation.} DA population difference $p_D(t)-p_A(t)$ as predicted by Eqs.~(\ref{me}) and~(\ref{Ldiss}) for different odorant dissipation rates: $\gamma_0=0$ (red dotted curves), $\gamma_0 = 1~\rm ns^{-1}$ (red dashed curves), and $\gamma_0 = 1000~\rm ns^{-1}$ (red solid curves). 
Exponential ET dynamics governed by the rates given in Eq.~(\ref{eq:Ahsansrate}), without mode dissipation, 
are also shown (points) along with the dynamics predicted by MJ theory (blue squares) for comparison. 
The bath cut-off frequency $\omega_c$ increases from top to bottom, with $\omega_c=k_BT/10$ in panel (i), $\omega_c=k_BT$ in panel (ii) and $\omega_c=2k_BT$ in panel (iii). The reorganisation energy is $\lambda = 15$~meV in each case and the odorant frequency is taken to be on resonance with the DA splitting, $\omega_0=\epsilon_D-\epsilon_A=200$~meV.}\label{dissipation}
\end{figure}

Given that, when present, the odorant is considered to be in the vicinity of the DA pair, it is reasonable to assume that it too interacts with the surrounding environment. 
Taking a linear coupling form we can model the effects of the resulting mode dissipation in a straightforward manner, once Born-Markov and rotating wave approximations are made, 
by introducing an additional Lindblad term to the right-hand-side of our master equation~(\ref{me}):~\cite{breuer_theory_2002}
\begin{align}
\mathcal{L}_{diss}[\tilde{\rho}_{\rm SP}]&=-\frac{\gamma_0}{2}N_0\big(a a^\dagger\tilde{\rho}_{\rm SP}-2a^\dagger\tilde{\rho}_{\rm SP}a+\tilde{\rho}_{\rm SP}a a^\dagger\big)\nonumber\\
&-\frac{\gamma_0}{2}(N_0+1)\big(a^\dagger a\tilde{\rho}_{\rm SP}-2a\tilde{\rho}_{\rm SP}a^\dagger+\tilde{\rho}_{\rm SP}a^\dagger a\big).\label{Ldiss}
\end{align}
Here, $\gamma_0$ is the odorant dissipation rate and $N_0=(e^{\omega_0/k_BT}-1)^{-1}$ is the 
phonon occupation number. 

In Fig.~\ref{dissipation} 
we investigate the effect of adding dissipation to the odorant. Specifically, we show a comparison of the dynamics in the absence of dissipation ($\gamma_0 = 0$, dotted curves), for moderate dissipation ($\gamma_0 \sim \Gamma_{D0A1}$, dashed curves) and for strong dissipation ($\gamma_0 \gg \Gamma_{D0A1}$, solid curves). In the limit of large dissipation, we see that the behaviour of the DA populations is consistent with that given by the transfer rates of Eq.~(\ref{eq:Ahsansrate}), and the description of the dynamics as an exponential ET process from donor to acceptor once again becomes valid (note the agreement between the points and solid curves). Furthermore, the dynamics only agrees with that predicted by MJ theory (blue squares) in the limit of low cut-off frequency in the bath (top panel). Importantly, we can also see that adding mode dissipation actually {\it assists} the transfer of population from donor to acceptor. In fact, the timescale for complete ET is substantially reduced as the dissipation rate is increased, thus ensuring a greater variance between the transfer time in the presence and absence of the odorant. 
In this manner, mode dissipation is seen to be beneficial for the molecular switching process, allowing easier discrimination in the receptor between the cases with and without the odorant present.

\subsection{Frequency resolution at strong dissipation}

\begin{figure}[h]
\includegraphics[width=0.35\textwidth]{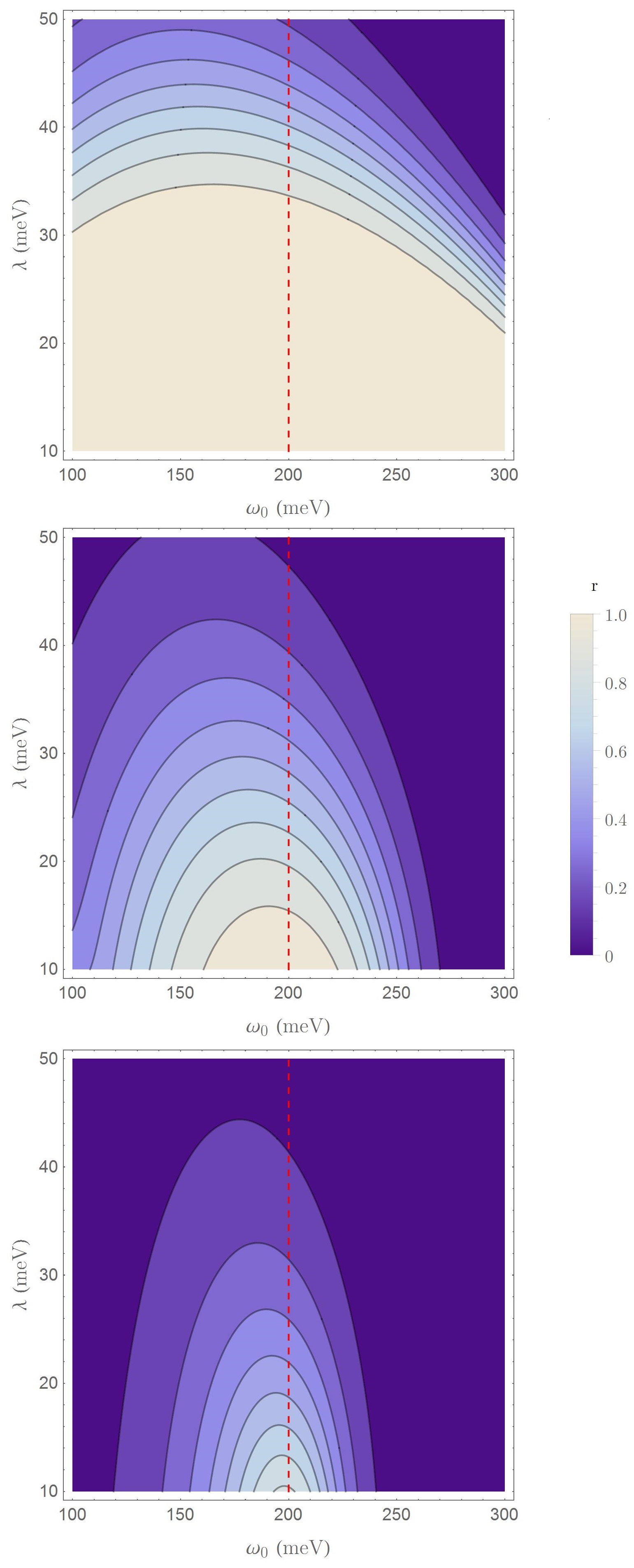}
\caption{{\bf Switch frequency resolution.} Contour plots of $r=\frac{(\Gamma_{\rm tot}-\Gamma_0)}{(\Gamma_{\rm tot}+\Gamma_0)}$ as a function of odorant frequency $\omega_0$ and bath reorganisation energy $\lambda$. 
Top: $\omega_c = k_B T /10$ (corresponding to the MJ limit). Middle: $\omega_c = k_B T$. Bottom: $\omega_c = 2 k_B T$. Other parameters are given in Table~\ref{partable}. 
Light regions correspond to $r\rightarrow1$ (good odorant discrimination) 
and the darkest regions correspond to $r\rightarrow0$ (poor discrimination). 
The red dashed lines shows the DA energy splitting.}\label{outofres}
\end{figure}

We can think of the olfactory model outlined 
herein in terms of a vibrational spectrometer, and one of the principle figures of merit for any spectroscopic device is its frequency resolution. The more 
sensitive the individual receptors are to resonance conditions between the DA pair and odorant, the better they will be able to distinguish the vibrational spectra of different molecules. To explore this aspect in our model, in Fig.~\ref{outofres} we map out the ratio 
$r=(\Gamma_{\rm tot}-\Gamma_0)/(\Gamma_{\rm tot}+\Gamma_0)$ of tunneling rates as a function of odorant frequency $\omega_0$ and bath reorganisation energy $\lambda$. 
Here, the definition of $\Gamma_{\rm tot}$ as the tunneling rate in the presence of the odorant and $\Gamma_0$ as the rate in its absence is made possible by the strong mode dissipation, $\gamma_0=1000$~ns$^{-1}$, which ensures exponential ET from donor to acceptor as just demonstrated. 
The top panel corresponds to the regime in which $\omega_c\ll k_B T$, therefore representing the behaviour predicted by the MJ rates of Eqs.~(\ref{daodorantabsent}) and~(\ref{daodorantpresent}), while the other panels encompass parameter ranges outside the MJ limit.

We see that in each plot there clearly exists a frequency window 
for which the tunneling rate is enhanced in the presence of the odorant above the background. Furthermore, this window 
becomes narrower as the environmental cut-off frequency is increased (middle and bottom panels), suggesting that the structure of the environment can play an important role in tuning the selectivity, and hence frequency resolution, of the receptor DA pair. 
Crucially, this cannot be predicted by the semiclassical MJ rates, 
which do not depend on the environmental cut-off, and predict a significant enhancement of tunnelling 
even for odorant vibrational frequencies relatively far off-resonant with the DA energy gap of $200$~meV (top panel). The low-frequency environment regime analysed in previous works---to which we have shown the MJ rates to apply under 
strong mode dissipation---is thus very sensitive the presence or absence of the odorant, but not necessarily to the odorant's specific vibrational spectrum. 

\begin{figure}[ht]
\centering
\includegraphics[width=0.45\textwidth]{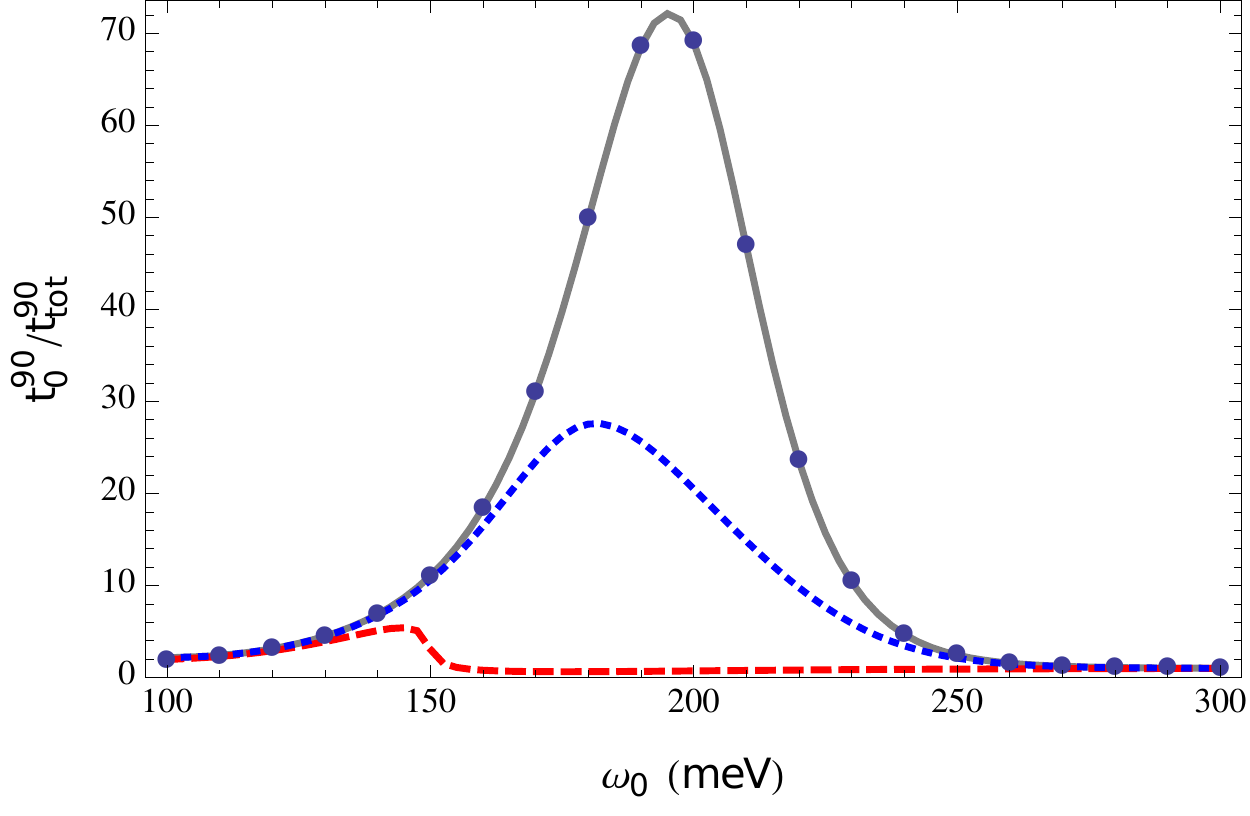}
\caption{{\bf Dissipation-assisted electron transfer.} The ratio of the electron transfer time (to $90\%$ acceptor population) in the absence of the odorant ($t^{90}_0$) to that in its presence ($t^{90}_{\rm tot}$) as a function of odorant frequency, as predicted by Eqs.~(\ref{me}) and~(\ref{Ldiss}). We take $\lambda = 10$~meV, $\omega_c = k_B T$, and all other parameters are from Table \ref{partable}. The three curves correspond to different levels of dissipation on the odorant: $\gamma_0 = 0$ (red, dashed), $\gamma_0 = 1~\rm ns^{-1}$ (blue, dotted) and $\gamma_0 = 1000~\rm ns^{-1}$ (grey, solid). The blue points correspond to the same ratio calculated using Eq.~(\ref{eq:Ahsansrate}), considering $n=0,1$.}
\label{peakplot}
\end{figure} 

It is evident from Fig.~\ref{outofres}, however, that increased frequency resolution comes at a cost. In order to achieve the more selective behaviour apparent 
with a larger environmental cut-off (and strong odorant dissipation), the coupling to the continuum environment 
must be reduced. 
Furthermore, the switch suffers from intrinsically higher levels of noise, since the peak values of the ratio $r$ also become smaller as the cut-off is increased. We can therefore conclude from our 
analysis that a tradeoff exists in which the combination of moderate to high frequency components in the environment and a lower reorganisation energy is advantageous for odorant vibrational frequency resolution, though not necessarily for discrimination simply between the presence and the absence of the odorant. It is worth noting also that the validity of the results derived from our polaron transformed Hamiltonian rely on the assumption $\lambda\gg V$. It is not possible, therefore, to extrapolate 
Fig.~\ref{outofres} 
to yet lower reorganisation energies within the present analysis. 

Finally, we stress that if we move out of the regime of strong odorant dissipation, 
we find that both the frequency resolution of the receptor and its sensitivity to the presence or absence of the odorant decreases. This is illustrated in Fig. \ref{peakplot}, which shows the ratio of the transfer time (to $90\%$ population on the acceptor site) in the absence of the odorant to that in its presence for different levels of odorant dissipation. We use the transfer time to allow a meaningful comparison here, since there is no single rate that accurately describes the dynamics outside the strongly dissipative case. As the dissipation rate is decreased, the height of the peak also decreases, while its width remains similar (dotted curve) until the dissipation ceases to play a role (dashed curve). Thus, as the mode dissipation is weakened, the receptor frequency resolution is severely compromised, 
as well as its capability to distinguish between the cases with and without the odorant present. 

\section{Summary}
\label{sec:conc}

We have developed a dynamical theory of a molecular switch and applied it to investigate vibrationally-assisted ET in the context of a proposed olfactory process. We have shown that the dynamics of the olfactory receptor can differ dramatically from that predicted by semiclassical MJ theory, 
even for a low-frequency environment, if vibrational dissipation is weak. 
The ET dynamics can, however, be brought into agreement with 
a simpler rate analysis---though not generally the MJ rates unless the semiclassical limit 
also applies---provided that strong dissipation is allowed to act on the odorant vibrational mode. This results in an enhanced switching with respect to the dissipationless case. Furthermore, we have found that low frequency environments, to which the MJ 
rates apply under strong odorant dissipation, do not provide good odorant frequency resolution in the ET rates. By modifying environmental parameters to move beyond the semiclassical limit it is, nevertheless, possible to substantially increase this resolution and thus select for odorants of a particular frequency.

While our 
present model is motivated by the problem of describing olfaction in biological systems, it actually corresponds to a wide variety of physical settings in which vibrationally-assisted transport processes are important. Examples include nano-mechanical oscillators coupled to two-level systems,~\cite{remus_damping_2012} such as quantum dots~\cite{lambert08} and superconducting qubits,~\cite{armour_probing_2008,armour_probing_2008b} as well as other biological systems such as the photosynthetic reaction centre in certain organisms.~\cite{xu94,blankenship_molecular_2002}   
Our results may also be especially relevant for the development of artificial molecular sensors, which could use the principles we have described 
in their design to aid in distinguishing chemical species based on their vibrational spectra.~\cite{lee_mimicking_2012,lee_bioelectronic_2012,du_recent_2013}

\section{Acknowledgements}
A.~C.~would like to thank V.~Vedral for fruitful discussions. A.~N.~is supported by Imperial College London and The University of Manchester. F.~A.~P.~is supported by the Leverhulme Trust. A.~C.~was supported by the National Research Foundation and Ministry of Education, Singapore.

\appendix
\section*{Appendix: Derivation of the master equation}\label{app:me}

In this Appendix we outline the general steps required to derive our polaron master equation. We start the analysis by 
writing the polaron-transformed Hamiltonian as (see Sec.~III) 
\begin{equation}
H_{\rm P}=U^{\dagger}HU=H_{\rm SP}+H_B+H_{\rm IP},
\end{equation}
and moving into the interaction picture with respect to $H_{\rm SP}$ and $H_B$. To derive the master equation we focus on the interaction part of the Hamiltonian which takes the form
\begin{align}
\tilde{H}_{\rm IP}(t)=&V\big(\ket{D}\bra{A}(t)\mathcal{A}_+(t)\mathcal{B}_+(t)
\nonumber\\
&+\ket{A}\bra{D}(t)\mathcal{A}_-(t)\mathcal{B}_-(t)\big)\nonumber\\
=&\sum_jS_j(t)\mathcal{B}_j(t),
\end{align}
with $j\,=\,\{+,-\}$ and 
\begin{align}
S_+(t)&=\,V|D\rangle\langle A|e^{i\epsilon t}\mathcal{A}_+(t),\nonumber\\
S_-(t)&=\,V|A\rangle\langle D|e^{-i\epsilon t}\mathcal{A}_-(t).
\end{align}
By considering the Liouville-von Neumann equation of motion for the density matrix of the whole system in the interaction picture, $\tilde{\chi}_{\rm P}(t)$, and tracing over the environment, we find:
\begin{align}
\frac{d}{dt}\tilde{\rho}_{\rm SP}(t) &= -i\int_0^t{\rm d}s{\rm tr}_B\left\{ \left[\tilde{H}_{\rm IP}(t),\left[\tilde{H}_{\rm IP}(s),\tilde{\chi}_{\rm P}(s)\right]\right]\right\}.\nonumber\\ 
 \label{eq:partialme}
\end{align}
Here, we have substituted in the solution to the full Liouville-von Neumann equation on the right-hand-side and have used ${\rm tr}\{[\tilde{H}_{\rm IP}(t),\chi_{\rm P}(0)]\}=0$. 
We then make the standard Born-Markov approximations. We assume that the evolution of the system is time local, such that it depends only on its current state, and that in the transformed frame the effective interaction `strength' is small, so that we may write $\tilde{\chi}_{\rm P}(t) \simeq \tilde{\rho}_{\rm SP}(t)\otimes \rho_{B}$ at all times, where $\rho_{B}$ is a thermal state of the environment. 
The Markov approximation amounts to replacing the state at time $s$ in the integrand of Eq.~(\ref{eq:partialme}) with the state at time $t$, changing variables such that $s\rightarrow t-\tau$, and taking the integral to infinity:
\begin{align}
\frac{d}{dt}\tilde{\rho}_{\rm SP}(t) =&-i \int_0^\infty{ {\rm d}\tau}\nonumber \\ 
&\times{\rm tr}_B\left\{\left[\tilde{H}_{\rm IP}(t),\left[\tilde{H}_{\rm IP}(t-\tau),\tilde{\rho}_{\rm SP}(t) \rho_{B}\right]\right]\right\}.
\end{align}
In terms of the operators $S_{\pm}(t)$ this now reads
\begin{align}
\frac{d}{dt}\tilde{\rho}_{\rm SP}(t)=&-\sum_{j,j'}\int_0^\infty d\tau \nonumber \\
&\times \bigg([S_j(t),S_{j'}(t-\tau)\tilde{\rho}_{\rm SP}(t)]C_{jj'}(\tau)\nonumber\\
&+[\tilde{\rho}_{\rm SP}(t)S_{j'}(t-\tau),S_j(t)]C_{j'j}(-\tau)\bigg),
\end{align}
where
\begin{equation}
C_{jj'}(t)\,=\,\tr_B\big(\mathcal{B}_j(\tau)\mathcal{B}_{j'}(0)\rho_B\big),
\end{equation}
and we have used $[\rho_B,H_B]=0$. There are actually two types of correlation function, $C_{\pm\pm}(\tau)$ and $C_{\pm\mp}(\tau)$, but one of them vanishes ($C_{\pm\pm}(\tau)\to0$) and the other satisfies $C_{\pm\mp}(-\tau)=[C_{\pm\mp}(\tau)]^*$. We therefore arrive at our polaron master equation
\begin{align}
\frac{d}{dt}\tilde{\rho}_{\rm SP}(t)=&-\int_0^\infty d\tau \nonumber \\
&\times \bigg\{\big([S_+(t),S_-(t-\tau)\tilde{\rho}_{\rm SP}(t)]+\nonumber\\
&+[S_-(t),S_+(t-\tau)\tilde{\rho}_{\rm SP}(t)]\big)C(\tau)+{\rm H.c.}\bigg\},
\end{align}
where we have written $C(\tau)\equiv C_{\pm\mp}(\tau)$. This is equivalent to Eq. (9) in the main text.


\begin{thebibliography}{73}%
\makeatletter
\providecommand \@ifxundefined [1]{%
 \@ifx{#1\undefined}
}%
\providecommand \@ifnum [1]{%
 \ifnum #1\expandafter \@firstoftwo
 \else \expandafter \@secondoftwo
 \fi
}%
\providecommand \@ifx [1]{%
 \ifx #1\expandafter \@firstoftwo
 \else \expandafter \@secondoftwo
 \fi
}%
\providecommand \natexlab [1]{#1}%
\providecommand \enquote  [1]{``#1''}%
\providecommand \bibnamefont  [1]{#1}%
\providecommand \bibfnamefont [1]{#1}%
\providecommand \citenamefont [1]{#1}%
\providecommand \href@noop [0]{\@secondoftwo}%
\providecommand \href [0]{\begingroup \@sanitize@url \@href}%
\providecommand \@href[1]{\@@startlink{#1}\@@href}%
\providecommand \@@href[1]{\endgroup#1\@@endlink}%
\providecommand \@sanitize@url [0]{\catcode `\\12\catcode `\$12\catcode
  `\&12\catcode `\#12\catcode `\^12\catcode `\_12\catcode `\%12\relax}%
\providecommand \@@startlink[1]{}%
\providecommand \@@endlink[0]{}%
\providecommand \url  [0]{\begingroup\@sanitize@url \@url }%
\providecommand \@url [1]{\endgroup\@href {#1}{\urlprefix }}%
\providecommand \urlprefix  [0]{URL }%
\providecommand \Eprint [0]{\href }%
\providecommand \doibase [0]{http://dx.doi.org/}%
\providecommand \selectlanguage [0]{\@gobble}%
\providecommand \bibinfo  [0]{\@secondoftwo}%
\providecommand \bibfield  [0]{\@secondoftwo}%
\providecommand \translation [1]{[#1]}%
\providecommand \BibitemOpen [0]{}%
\providecommand \bibitemStop [0]{}%
\providecommand \bibitemNoStop [0]{.\EOS\space}%
\providecommand \EOS [0]{\spacefactor3000\relax}%
\providecommand \BibitemShut  [1]{\csname bibitem#1\endcsname}%
\let\auto@bib@innerbib\@empty
%</preamble>
\bibitem [{\citenamefont {Nitzan}(2006)}]{nitzan_chemical_2006}%
  \BibitemOpen
  \bibfield  {author} {\bibinfo {author} {\bibfnamefont {A.}~\bibnamefont
  {Nitzan}},\ }\href@noop {} {{\selectlanguage {english}\emph {\bibinfo {title}
  {Chemical Dynamics in Condensed Phases: Relaxation, Transfer and Reactions in
  Condensed Molecular Systems}}}}\ (\bibinfo
  {publisher} {Oxford University Press},\ \bibinfo {year} {2006})\BibitemShut {NoStop}%
\bibitem [{\citenamefont {May}\ and\ \citenamefont
  {K\"{u}hn}(2011)}]{may_kuhn_book}%
  \BibitemOpen
  \bibfield  {author} {\bibinfo {author} {\bibfnamefont {V.}~\bibnamefont
  {May}}\ and\ \bibinfo {author} {\bibfnamefont {O.}~\bibnamefont {K\"{u}hn}},\
  }\href@noop {} {{\selectlanguage {english}\emph {\bibinfo {title} {Charge and
  Energy Transfer Dynamics in Molecular Systems}}}}\ (\bibinfo  {publisher}
  {Wiley},\ \bibinfo {year} {2011})\BibitemShut {NoStop}%
\bibitem [{\citenamefont {Lambert}\ \emph {et~al.}(2013)\citenamefont
  {Lambert}, \citenamefont {Chen}, \citenamefont {Cheng}, \citenamefont {Li},
  \citenamefont {Chen},\ and\ \citenamefont {Nori}}]{lambert_review13}%
  \BibitemOpen
  \bibfield  {author} {\bibinfo {author} {\bibfnamefont {N.}~\bibnamefont
  {Lambert}}, \bibinfo {author} {\bibfnamefont {Y.-N.}\ \bibnamefont {Chen}},
  \bibinfo {author} {\bibfnamefont {Y.-C.}\ \bibnamefont {Cheng}}, \bibinfo
  {author} {\bibfnamefont {C.-M.}\ \bibnamefont {Li}}, \bibinfo {author}
  {\bibfnamefont {G.-Y.}\ \bibnamefont {Chen}}, \ and\ \bibinfo {author}
  {\bibfnamefont {F.}~\bibnamefont {Nori}},\ }\href@noop {} %{{\selectlanguage {english}\emph {\bibinfo {title}{Quantum biology,}}}}
  \ {\bibfield
  {journal} {\bibinfo  {journal} {Nature Phys.}\ }\textbf {\bibinfo {volume}
  {9}},\ \bibinfo {pages} {10} (\bibinfo {year} {2013})}\BibitemShut {NoStop}%
\bibitem [{\citenamefont {Cheng}\ and\ \citenamefont
  {Fleming}(2009)}]{cheng09}%
  \BibitemOpen
  \bibfield  {author} {\bibinfo {author} {\bibfnamefont {Y.-C.}\ \bibnamefont
  {Cheng}}\ and\ \bibinfo {author} {\bibfnamefont {G.~R.}\ \bibnamefont
  {Fleming}},\ }\href@noop {} %{{\selectlanguage {english}\emph {\bibinfo {title}  {Dynamics of light harvesting in photosynthesis,}}}}
\ {\bibfield  {journal} {\bibinfo  {journal} {Annu.
  Rev. Phys. Chem.}\ }\textbf {\bibinfo {volume} {60}},\ \bibinfo {pages} {241}
  (\bibinfo {year} {2009})}\BibitemShut {NoStop}%
\bibitem [{\citenamefont {Olaya-Castro}\ and\ \citenamefont
  {Scholes}(2011)}]{olaya_castro11}%
  \BibitemOpen
  \bibfield  {author} {\bibinfo {author} {\bibfnamefont {A.}~\bibnamefont
  {Olaya-Castro}}\ and\ \bibinfo {author} {\bibfnamefont {G.~D.}\ \bibnamefont
  {Scholes}},\ }\href@noop {} %{{\selectlanguage {english}\emph {\bibinfo {title}  {Energy transfer from Forster-Dexter theory to quantum coherent light-harvesting,}}}}
\ {\bibfield  {journal} {\bibinfo  {journal} {Int.
  Rev. Phys. Chem.}\ }\textbf {\bibinfo {volume} {30}},\ \bibinfo {pages} {49}
  (\bibinfo {year} {2011})}\BibitemShut {NoStop}%
\bibitem [{\citenamefont {Huelga}\ and\ \citenamefont
  {Plenio}(2011)}]{huelga11}%
  \BibitemOpen
  \bibfield  {author} {\bibinfo {author} {\bibfnamefont {S.~F.}\ \bibnamefont
  {Huelga}}\ and\ \bibinfo {author} {\bibfnamefont {M.~B.}\ \bibnamefont
  {Plenio}},\ }\href@noop {} %{{\selectlanguage {english}\emph {\bibinfo {title}  {Quantum dynamics of bio-molecular systems in noisy environments,}}}}
\ {\bibfield  {journal} {\bibinfo  {journal}
  {Procedia Chem.}\ }\textbf {\bibinfo {volume} {3}},\ \bibinfo {pages} {248}
  (\bibinfo {year} {2011})}\BibitemShut {NoStop}%
\bibitem [{\citenamefont {Breuer}\ and\ \citenamefont
  {Petruccione}(2002)}]{breuer_theory_2002}%
  \BibitemOpen
  \bibfield  {author} {\bibinfo {author} {\bibfnamefont {H.-P.}\ \bibnamefont
  {Breuer}}\ and\ \bibinfo {author} {\bibfnamefont {F.}~\bibnamefont
  {Petruccione}},\ }\href@noop {} {{\selectlanguage {english}\emph {\bibinfo
  {title} {The theory of open quantum systems}}}}\ (\bibinfo  {publisher}
  {Oxford University Press},\ \bibinfo
  {year} {2002})\BibitemShut {NoStop}%
\bibitem [{\citenamefont {Ishizaki}\ \emph {et~al.}(2010)\citenamefont
  {Ishizaki}, \citenamefont {Calhoun}, \citenamefont {Schlau-Cohen},\ and\
  \citenamefont {Fleming}}]{ishizakireview10}%
  \BibitemOpen
  \bibfield  {author} {\bibinfo {author} {\bibfnamefont {A.}~\bibnamefont
  {Ishizaki}}, \bibinfo {author} {\bibfnamefont {T.~R.}\ \bibnamefont
  {Calhoun}}, \bibinfo {author} {\bibfnamefont {G.~S.}\ \bibnamefont
  {Schlau-Cohen}}, \ and\ \bibinfo {author} {\bibfnamefont {G.~R.}\
  \bibnamefont {Fleming}},\ }\href@noop {} %{{\selectlanguage {english}\emph {\bibinfo {title}  {Quantum coherence and its interplay with protein environments in photosynthetic electronic energy transfer,}}}}
\ {\bibfield  {journal} {\bibinfo
  {journal} {Phys. Chem. Chem. Phys.}\ }\textbf {\bibinfo {volume} {12}},\
  \bibinfo {pages} {7319} (\bibinfo {year} {2010})}\BibitemShut {NoStop}%
\bibitem [{\citenamefont {Garg}\ \emph {et~al.}(1985)\citenamefont {Garg},
  \citenamefont {Onuchic},\ and\ \citenamefont {Ambegaokar}}]{garg85}%
  \BibitemOpen
  \bibfield  {author} {\bibinfo {author} {\bibfnamefont {A.}~\bibnamefont
  {Garg}}, \bibinfo {author} {\bibfnamefont {J.~N.}\ \bibnamefont {Onuchic}}, \
  and\ \bibinfo {author} {\bibfnamefont {V.}~\bibnamefont {Ambegaokar}},\
  }\href@noop {} %{{\selectlanguage {english}\emph {\bibinfo {title}  {Effect of friction on electron transfer in biomolecules,}}}}
\ {\bibfield  {journal} {\bibinfo  {journal} {J. Chem. Phys.}\
  }\textbf {\bibinfo {volume} {83}},\ \bibinfo {pages} {4491} (\bibinfo {year}
  {1985})}\BibitemShut {NoStop}%
\bibitem [{\citenamefont {Olbrich}\ \emph {et~al.}(2011)\citenamefont
  {Olbrich}, \citenamefont {Str{\"u}mpfer}, \citenamefont {Schulten},\ and\
  \citenamefont {Kleinekath{\"o}fer}}]{olbrich11}%
  \BibitemOpen
  \bibfield  {author} {\bibinfo {author} {\bibfnamefont {C.}~\bibnamefont
  {Olbrich}}, \bibinfo {author} {\bibfnamefont {J.}~\bibnamefont
  {Str{\"u}mpfer}}, \bibinfo {author} {\bibfnamefont {K.}~\bibnamefont
  {Schulten}}, \ and\ \bibinfo {author} {\bibfnamefont {U.}~\bibnamefont
  {Kleinekath{\"o}fer}},\ }\href@noop {} %{{\selectlanguage {english}\emph {\bibinfo {title}  {Theory and Simulation of the Environmental Effects on FMO Electronic Transitions,}}}}
\ {\bibfield  {journal} {\bibinfo
  {journal} {J. Phys. Chem. Lett.}\ }\textbf {\bibinfo {volume} {2}},\ \bibinfo
  {pages} {1771} (\bibinfo {year} {2011})}\BibitemShut {NoStop}%
\bibitem [{\citenamefont {Shim}\ \emph {et~al.}(2012)\citenamefont {Shim},
  \citenamefont {Rebentrost}, \citenamefont {Valleau},\ and\ \citenamefont
  {Aspuru-Guzik}}]{shim12}%
  \BibitemOpen
  \bibfield  {author} {\bibinfo {author} {\bibfnamefont {S.}~\bibnamefont
  {Shim}}, \bibinfo {author} {\bibfnamefont {P.}~\bibnamefont {Rebentrost}},
  \bibinfo {author} {\bibfnamefont {S.}~\bibnamefont {Valleau}}, \ and\
  \bibinfo {author} {\bibfnamefont {A.}~\bibnamefont {Aspuru-Guzik}},\
  }\href@noop {} %{{\selectlanguage {english}\emph {\bibinfo {title}  {Atomistic Study of the Long-Lived Quantum Coherences in the Fenna-Matthews-Olson Complex,}}}}
\ {\bibfield  {journal} {\bibinfo  {journal} {Biophys. J.}\
  }\textbf {\bibinfo {volume} {102}},\ \bibinfo {pages} {649} (\bibinfo {year}
  {2012})}\BibitemShut {NoStop}%
\bibitem [{\citenamefont {Rey}\ \emph {et~al.}(2013)\citenamefont {Rey},
  \citenamefont {Chin}, \citenamefont {Huelga},\ and\ \citenamefont
  {Plenio}}]{phonon_antenna2013}%
  \BibitemOpen
  \bibfield  {author} {\bibinfo {author} {\bibfnamefont {M.~d.}\ \bibnamefont
  {Rey}}, \bibinfo {author} {\bibfnamefont {A.~W.}\ \bibnamefont {Chin}},
  \bibinfo {author} {\bibfnamefont {S.~F.}\ \bibnamefont {Huelga}}, \ and\
  \bibinfo {author} {\bibfnamefont {M.~B.}\ \bibnamefont {Plenio}},\
  }\href@noop {} %{{\selectlanguage {english}\emph {\bibinfo {title}  {Exploiting Structured Environments for Efficient Energy Transfer: The Phonon Antenna Mechanism,}}}}
\ {\bibfield  {journal} {\bibinfo  {journal} {J. Phys. Chem.
  Lett.}\ }\textbf {\bibinfo {volume} {4}},\ \bibinfo {pages} {903} (\bibinfo
  {year} {2013})}\BibitemShut {NoStop}%
\bibitem [{\citenamefont {Chin}\ \emph {et~al.}(2013)\citenamefont {Chin},
  \citenamefont {Prior}, \citenamefont {Rosenbach}, \citenamefont
  {Caycedo-Soler}, \citenamefont {Huelga},\ and\ \citenamefont
  {Plenio}}]{chin_role_2013}%
  \BibitemOpen
  \bibfield  {author} {\bibinfo {author} {\bibfnamefont {A.~W.}\ \bibnamefont
  {Chin}}, \bibinfo {author} {\bibfnamefont {J.}~\bibnamefont {Prior}},
  \bibinfo {author} {\bibfnamefont {R.}~\bibnamefont {Rosenbach}}, \bibinfo
  {author} {\bibfnamefont {F.}~\bibnamefont {Caycedo-Soler}}, \bibinfo {author}
  {\bibfnamefont {S.~F.}\ \bibnamefont {Huelga}}, \ and\ \bibinfo {author}
  {\bibfnamefont {M.~B.}\ \bibnamefont {Plenio}},\ }\href@noop {} %{{\selectlanguage {english}\emph {\bibinfo {title}  {The role of non-equilibrium vibrational structures in electronic coherence and recoherence in pigment-protein complexes,}}}}
\ {\bibfield  {journal} {\bibinfo  {journal} {Nature Phys.}\ }\textbf {\bibinfo
  {volume} {9}},\ \bibinfo {pages} {113} (\bibinfo {year} {2013})}\BibitemShut
  {NoStop}%
\bibitem [{\citenamefont {{O'Reilly}}\ and\ \citenamefont
  {Olaya-Castro}(2014)}]{oreilly_non-classicality_2014}%
  \BibitemOpen
  \bibfield  {author} {\bibinfo {author} {\bibfnamefont {E.~J.}\ \bibnamefont
  {{O'Reilly}}}\ and\ \bibinfo {author} {\bibfnamefont {A.}~\bibnamefont
  {Olaya-Castro}},\ }\href@noop {} %{{\selectlanguage {english}\emph {\bibinfo {title}  {Non-classicality of the molecular vibrations assisting exciton energy transfer at room temperature,}}}}
\ {\bibfield
  {journal} {\bibinfo  {journal} {Nature Commun.}\ }\textbf {\bibinfo {volume}
  {5}}, \ \bibinfo {pages} {3012} (\bibinfo {year} {2014})}\BibitemShut {NoStop}%
\bibitem [{\citenamefont {{Irish}}\ \emph {et~al.}()\citenamefont {{Irish}},
  \citenamefont {{G{\'o}mez-Bombarelli}},\ and\ \citenamefont
  {{Lovett}}}]{Elinor_Irish2013}%
  \BibitemOpen
  \bibfield  {author} {\bibinfo {author} {\bibfnamefont {E.~K.}\ \bibnamefont
  {{Irish}}}, \bibinfo {author} {\bibfnamefont {R.}~\bibnamefont
  {{G{\'o}mez-Bombarelli}}}, \ and\ \bibinfo {author} {\bibfnamefont {B.~W.}\
  \bibnamefont {{Lovett}}},\ }\href@noop {} %{{\selectlanguage {english}\emph {\bibinfo {title}  {Vibration-assisted resonance in photosynthetic excitation energy transfer,}}}}
\  {\bibfield  {journal} {\bibinfo  {journal} {Phys. Rev. A}\ }\textbf
  {\bibinfo {volume} {90}},\ \bibinfo {pages} {012510} (\bibinfo {year}
  {2014})}\BibitemShut {NoStop}%
\bibitem [{\citenamefont {Ritschel}\ \emph {et~al.}(2011)\citenamefont
  {Ritschel}, \citenamefont {Roden}, \citenamefont {Strunz},\ and\
  \citenamefont {Eisfeld}}]{ritschel11}%
  \BibitemOpen
  \bibfield  {author} {\bibinfo {author} {\bibfnamefont {G.}~\bibnamefont
  {Ritschel}}, \bibinfo {author} {\bibfnamefont {J.}~\bibnamefont {Roden}},
  \bibinfo {author} {\bibfnamefont {W.~T.}\ \bibnamefont {Strunz}}, \ and\
  \bibinfo {author} {\bibfnamefont {A.}~\bibnamefont {Eisfeld}},\ }\href@noop
  {} %{{\selectlanguage {english}\emph {\bibinfo {title}  {An efficient method to calculate excitation energy transfer in light-harvesting systems: application to the Fenna-Matthews-Olson complex,}}}}
\  {\bibfield  {journal} {\bibinfo  {journal} {New J. Phys.}\ }\textbf
  {\bibinfo {volume} {13}},\ \bibinfo {pages} {113034} (\bibinfo {year}
  {2011})}\BibitemShut {NoStop}%
\bibitem [{\citenamefont {Lim}\ \emph {et~al.}()\citenamefont {Lim},
  \citenamefont {Tame}, \citenamefont {Yee}, \citenamefont {Lee},\ and\
  \citenamefont {Lee}}]{lim13}%
  \BibitemOpen
  \bibfield  {author} {\bibinfo {author} {\bibfnamefont {J.}~\bibnamefont
  {Lim}}, \bibinfo {author} {\bibfnamefont {M.}~\bibnamefont {Tame}}, \bibinfo
  {author} {\bibfnamefont {K.~H.}\ \bibnamefont {Yee}}, \bibinfo {author}
  {\bibfnamefont {J.-S.}\ \bibnamefont {Lee}}, \ and\ \bibinfo {author}
  {\bibfnamefont {J.}~\bibnamefont {Lee}},\ }\href@noop {} %{{\selectlanguage {english}\emph {\bibinfo {title}  {Phonon-induced dynamic resonance energy transfer,}}}}
\ {\bibfield  {journal} {\bibinfo  {journal} {New J. Phys.}\
  }\textbf {\bibinfo {volume} {16}},\ \bibinfo {pages} {053018} (\bibinfo
  {year} {2014})}\BibitemShut {NoStop}%
\bibitem [{\citenamefont {Kolli}\ \emph {et~al.}(2012)\citenamefont {Kolli},
  \citenamefont {O'Reilly}, \citenamefont {Scholes},\ and\ \citenamefont
  {Olaya-Castro}}]{kolli12}%
  \BibitemOpen
  \bibfield  {author} {\bibinfo {author} {\bibfnamefont {A.}~\bibnamefont
  {Kolli}}, \bibinfo {author} {\bibfnamefont {E.~J.}\ \bibnamefont {O'Reilly}},
  \bibinfo {author} {\bibfnamefont {G.~D.}\ \bibnamefont {Scholes}}, \ and\
  \bibinfo {author} {\bibfnamefont {A.}~\bibnamefont {Olaya-Castro}},\
  }\href@noop {} %{{\selectlanguage {english}\emph {\bibinfo {title}  {The fundamental role of quantized vibrations in coherent light harvesting by cryptophyte algae,}}}}
\ {\bibfield  {journal} {\bibinfo  {journal} {J. Chem. Phys.}\
  }\textbf {\bibinfo {volume} {137}},\ \bibinfo {pages} {174109} (\bibinfo
  {year} {2012})}\BibitemShut {NoStop}%
\bibitem [{\citenamefont {Christensson}\ \emph {et~al.}(2012)\citenamefont
  {Christensson}, \citenamefont {Kauffmann}, \citenamefont {Pullerits},\ and\
  \citenamefont {Man{\v{c}}al}}]{christensson12}%
  \BibitemOpen
  \bibfield  {author} {\bibinfo {author} {\bibfnamefont {N.}~\bibnamefont
  {Christensson}}, \bibinfo {author} {\bibfnamefont {H.~F.}\ \bibnamefont
  {Kauffmann}}, \bibinfo {author} {\bibfnamefont {T.}~\bibnamefont
  {Pullerits}}, \ and\ \bibinfo {author} {\bibfnamefont {T.}~\bibnamefont
  {Man{\v{c}}al}},\ }\href@noop {} %{{\selectlanguage {english}\emph {\bibinfo {title}  {Origin of Long-Lived Coherences in Light-Harvesting Complexes,}}}}
\ {\bibfield  {journal} {\bibinfo  {journal}
  {J. Phys. Chem. B}\ }\textbf {\bibinfo {volume} {116}},\ \bibinfo {pages}
  {7449} (\bibinfo {year} {2012})}\BibitemShut {NoStop}%
\bibitem [{\citenamefont {Chenu}\ \emph {et~al.}(2013)\citenamefont {Chenu},
  \citenamefont {Christensson}, \citenamefont {Kauffmann},\ and\ \citenamefont
  {Man{\v{c}}al}}]{chenu13}%
  \BibitemOpen
  \bibfield  {author} {\bibinfo {author} {\bibfnamefont {A.}~\bibnamefont
  {Chenu}}, \bibinfo {author} {\bibfnamefont {N.}~\bibnamefont {Christensson}},
  \bibinfo {author} {\bibfnamefont {H.~F.}\ \bibnamefont {Kauffmann}}, \ and\
  \bibinfo {author} {\bibfnamefont {T.}~\bibnamefont {Man{\v{c}}al}},\
  }\href@noop {} %{{\selectlanguage {english}\emph {\bibinfo {title}  {Enhancement of vibronic and ground-state vibrational coherences in 2D spectra of photosynthetic complexes,}}}}
\ {\bibfield  {journal} {\bibinfo  {journal} {Sci. Rep.}\
  }\textbf {\bibinfo {volume} {3}},\ \bibinfo {pages} {2029} (\bibinfo {year}
  {2013})}\BibitemShut {NoStop}%
\bibitem [{\citenamefont {Turin}(1996)}]{turin_spectroscopic_1996}%
  \BibitemOpen
  \bibfield  {author} {\bibinfo {author} {\bibfnamefont {L.}~\bibnamefont
  {Turin}},\ }\href@noop {} %{{\selectlanguage {english}\emph {\bibinfo {title}  {A Spectroscopic Mechanism for Primary Olfactory Reception,}}}}
\ {\bibfield  {journal} {\bibinfo  {journal} {Chem. Senses}\ }\textbf {\bibinfo
  {volume} {21}},\ \bibinfo {pages} {773} (\bibinfo {year} {1996})}\BibitemShut
  {NoStop}%
\bibitem [{\citenamefont {Turin}(2002)}]{turin_method_2002}%
  \BibitemOpen
  \bibfield  {author} {\bibinfo {author} {\bibfnamefont {L.}~\bibnamefont
  {Turin}},\ }\href@noop {} %{{\selectlanguage {english}\emph {\bibinfo {title}  {A Method for the Calculation of Odor Character from Molecular Structure,}}}}
\ {\bibfield  {journal} {\bibinfo  {journal} {J. Theor. Biol.}\ }\textbf
  {\bibinfo {volume} {216}},\ \bibinfo {pages} {367} (\bibinfo {year}
  {2002})}\BibitemShut {NoStop}%
\bibitem [{\citenamefont {Keller}\ and\ \citenamefont
  {Vosshall}(2004)}]{keller_psychophysical_2004}%
  \BibitemOpen
  \bibfield  {author} {\bibinfo {author} {\bibfnamefont {A.}~\bibnamefont
  {Keller}}\ and\ \bibinfo {author} {\bibfnamefont {L.~B.}\ \bibnamefont
  {Vosshall}},\ }\href@noop {} %{{\selectlanguage {english}\emph {\bibinfo {title}  {A psychophysical test of the vibration theory of olfaction,}}}}
\ {\bibfield
  {journal} {\bibinfo  {journal} {Nature Neurosci.}\ }\textbf {\bibinfo {volume}
  {7}},\ \bibinfo {pages} {337} (\bibinfo {year} {2004})}\BibitemShut {NoStop}%
\bibitem [{\citenamefont {Brookes}\ \emph {et~al.}(2007)\citenamefont
  {Brookes}, \citenamefont {Hartoutsiou}, \citenamefont {Horsfield},\ and\
  \citenamefont {Stoneham}}]{brookes_could_2007}%
  \BibitemOpen
  \bibfield  {author} {\bibinfo {author} {\bibfnamefont {J.~C.}\ \bibnamefont
  {Brookes}}, \bibinfo {author} {\bibfnamefont {F.}~\bibnamefont
  {Hartoutsiou}}, \bibinfo {author} {\bibfnamefont {A.~P.}\ \bibnamefont
  {Horsfield}}, \ and\ \bibinfo {author} {\bibfnamefont {A.~M.}\ \bibnamefont
  {Stoneham}},\ }\href@noop {} %{{\selectlanguage {english}\emph {\bibinfo {title}  {Could Humans Recognize Odor by Phonon Assisted Tunneling?,}}}}
\ {\bibfield  {journal} {\bibinfo  {journal} {Phys. Rev. Lett.}\ }\textbf
  {\bibinfo {volume} {98}},\ \bibinfo {pages} {038101} (\bibinfo {year}
  {2007})}\BibitemShut {NoStop}%
\bibitem [{\citenamefont {Hettinger}(2011)}]{hettinger_olfaction_2011}%
  \BibitemOpen
  \bibfield  {author} {\bibinfo {author} {\bibfnamefont {T.~P.}\ \bibnamefont
  {Hettinger}},\ }\href@noop {} %{{\selectlanguage {english}\emph {\bibinfo {title}  {Olfaction is a chemical sense, not a spectral sense,}}}}
\ {\bibfield
  {journal} {\bibinfo  {journal} {PNAS}\ }\textbf {\bibinfo {volume} {108}},\
  \bibinfo {pages} {E349} (\bibinfo {year} {2011})}\BibitemShut {NoStop}%
\bibitem [{\citenamefont {Franco}\ \emph {et~al.}(2011)\citenamefont {Franco},
  \citenamefont {Turin}, \citenamefont {Mershin},\ and\ \citenamefont
  {Skoulakis}}]{franco_molecular_2011}%
  \BibitemOpen
  \bibfield  {author} {\bibinfo {author} {\bibfnamefont {M.~I.}\ \bibnamefont
  {Franco}}, \bibinfo {author} {\bibfnamefont {L.}~\bibnamefont {Turin}},
  \bibinfo {author} {\bibfnamefont {A.}~\bibnamefont {Mershin}}, \ and\
  \bibinfo {author} {\bibfnamefont {E.~M.~C.}\ \bibnamefont {Skoulakis}},\
  }\href@noop {} %{{\selectlanguage {english}\emph {\bibinfo {title}  {Molecular vibration-sensing component in Drosophila melanogaster olfaction,}}}}
\ {\bibfield  {journal}
  {\bibinfo  {journal} {PNAS}\ }\textbf {\bibinfo {volume} {108}},\ \bibinfo
  {pages} {3797} (\bibinfo {year} {2011})}\BibitemShut {NoStop}%
\bibitem [{\citenamefont {Solov'yov}\ \emph {et~al.}(2012)\citenamefont
  {Solov'yov}, \citenamefont {Chang},\ and\ \citenamefont
  {Schulten}}]{solovyov_vibrationally_2012}%
  \BibitemOpen
  \bibfield  {author} {\bibinfo {author} {\bibfnamefont {I.~A.}\ \bibnamefont
  {Solov'yov}}, \bibinfo {author} {\bibfnamefont {P.-Y.}\ \bibnamefont
  {Chang}}, \ and\ \bibinfo {author} {\bibfnamefont {K.}~\bibnamefont
  {Schulten}},\ }\href@noop {} %{{\selectlanguage {english}\emph {\bibinfo {title}  {Vibrationally assisted electron transfer mechanism of olfaction: myth or reality?,}}}}
\ {\bibfield
   {journal} {\bibinfo  {journal} {Phys. Chem. Chem. Phys.}\ }\textbf {\bibinfo
  {volume} {14}},\ \bibinfo {pages} {13861} (\bibinfo {year}
  {2012})}\BibitemShut {NoStop}%
\bibitem [{\citenamefont {Kovacic}(2012)}]{kovacic_mechanism_2012}%
  \BibitemOpen
  \bibfield  {author} {\bibinfo {author} {\bibfnamefont {P.}~\bibnamefont
  {Kovacic}},\ }\href@noop {} %{{\selectlanguage {english}\emph {\bibinfo {title}  {Mechanism of smell: Electrochemistry, receptors and cell signaling,}}}}
\ {\bibfield  {journal} {\bibinfo  {journal} {J. Electrostat.}\ }\textbf
  {\bibinfo {volume} {70}},\ \bibinfo {pages} {1} (\bibinfo {year}
  {2012})}\BibitemShut {NoStop}%
\bibitem [{\citenamefont {Bittner}\ \emph {et~al.}(2012)\citenamefont
  {Bittner}, \citenamefont {Madalan}, \citenamefont {Czader},\ and\
  \citenamefont {Roman}}]{bittner_quantum_2012}%
  \BibitemOpen
  \bibfield  {author} {\bibinfo {author} {\bibfnamefont {E.~R.}\ \bibnamefont
  {Bittner}}, \bibinfo {author} {\bibfnamefont {A.}~\bibnamefont {Madalan}},
  \bibinfo {author} {\bibfnamefont {A.}~\bibnamefont {Czader}}, \ and\ \bibinfo
  {author} {\bibfnamefont {G.}~\bibnamefont {Roman}},\ }\href@noop {} %{{\selectlanguage {english}\emph {\bibinfo {title}  {Quantum origins of molecular recognition and olfaction in drosophila,}}}}
\ {\bibfield  {journal} {\bibinfo  {journal} {J. Chem. Phys.}\ }\textbf
  {\bibinfo {volume} {137}},\ \bibinfo {pages} {22A551} (\bibinfo {year}
  {2012})}\BibitemShut {NoStop}%
\bibitem [{\citenamefont {Gane}\ \emph {et~al.}(2013)\citenamefont {Gane},
  \citenamefont {Georganakis}, \citenamefont {Maniati}, \citenamefont
  {Vamvakias}, \citenamefont {Ragoussis}, \citenamefont {Skoulakis},\ and\
  \citenamefont {Turin}}]{gane_molecular_2013}%
  \BibitemOpen
  \bibfield  {author} {\bibinfo {author} {\bibfnamefont {S.}~\bibnamefont
  {Gane}}, \bibinfo {author} {\bibfnamefont {D.}~\bibnamefont {Georganakis}},
  \bibinfo {author} {\bibfnamefont {K.}~\bibnamefont {Maniati}}, \bibinfo
  {author} {\bibfnamefont {M.}~\bibnamefont {Vamvakias}}, \bibinfo {author}
  {\bibfnamefont {N.}~\bibnamefont {Ragoussis}}, \bibinfo {author}
  {\bibfnamefont {E.~M.~C.}\ \bibnamefont {Skoulakis}}, \ and\ \bibinfo
  {author} {\bibfnamefont {L.}~\bibnamefont {Turin}},\ }\href@noop {} %{{\selectlanguage {english}\emph {\bibinfo {title}  {Molecular Vibration-Sensing Component in Human Olfaction,}}}}
\ {\bibfield  {journal}
  {\bibinfo  {journal} {{PLoS} {ONE}}\ }\textbf {\bibinfo {volume} {8}},\
  \bibinfo {pages} {e55780} (\bibinfo {year} {2013})}\BibitemShut {NoStop}%
\bibitem [{\citenamefont {Rowe}(2005)}]{rowe_chemistry_2005}%
  \BibitemOpen
  \bibfield  {author} {\bibinfo {author} {\bibfnamefont {D.~J.}\ \bibnamefont
  {Rowe}},\ }\href@noop {} {{\selectlanguage {english}\emph {\bibinfo {title}
  {Chemistry and technology of flavors and fragrances}}}}\ (\bibinfo
  {publisher} {Blackwell},\ \bibinfo {address} {Oxford},\ \bibinfo {year} {2005})\BibitemShut {NoStop}%
\bibitem [{\citenamefont {Axel}(2005)}]{axel_scents_2005}%
  \BibitemOpen
  \bibfield  {author} {\bibinfo {author} {\bibfnamefont {R.}~\bibnamefont
  {Axel}},\ }\href@noop {} %{{\selectlanguage {english}\emph {\bibinfo {title}  {Scents and sensibility: a molecular logic of olfactory perception (Nobel lecture),}}}}
\ {\bibfield  {journal} {\bibinfo  {journal} {Angew. Chem. Int. Ed.}\ }\textbf
  {\bibinfo {volume} {44}},\ \bibinfo {pages} {6110} (\bibinfo {year}
  {2005})}\BibitemShut {NoStop}%
\bibitem [{\citenamefont {Buck}(2005)}]{buck_unraveling_2005}%
  \BibitemOpen
  \bibfield  {author} {\bibinfo {author} {\bibfnamefont {L.~B.}\ \bibnamefont
  {Buck}},\ }\href@noop {} %{{\selectlanguage {english}\emph {\bibinfo {title}  {Unraveling the Sense of Smell (Nobel Lecture),}}}}
\ {\bibfield  {journal} {\bibinfo  {journal} {Angew. Chem. Int. Ed.}\ }\textbf
  {\bibinfo {volume} {44}},\ \bibinfo {pages} {6128} (\bibinfo {year}
  {2005})}\BibitemShut {NoStop}%
\bibitem [{\citenamefont {Lee}\ \emph {et~al.}(2012{\natexlab{a}})\citenamefont
  {Lee}, \citenamefont {Kwon}, \citenamefont {Song}, \citenamefont {Park},
  \citenamefont {Sung}, \citenamefont {Jang},\ and\ \citenamefont
  {Park}}]{lee_mimicking_2012}%
  \BibitemOpen
  \bibfield  {author} {\bibinfo {author} {\bibfnamefont {S.~H.}\ \bibnamefont
  {Lee}}, \bibinfo {author} {\bibfnamefont {O.~S.}\ \bibnamefont {Kwon}},
  \bibinfo {author} {\bibfnamefont {H.~S.}\ \bibnamefont {Song}}, \bibinfo
  {author} {\bibfnamefont {S.~J.}\ \bibnamefont {Park}}, \bibinfo {author}
  {\bibfnamefont {J.~H.}\ \bibnamefont {Sung}}, \bibinfo {author}
  {\bibfnamefont {J.}~\bibnamefont {Jang}}, \ and\ \bibinfo {author}
  {\bibfnamefont {T.~H.}\ \bibnamefont {Park}},\ }\href@noop {} %{{\selectlanguage {english}\emph {\bibinfo {title}  {Mimicking the human smell sensing mechanism with an artificial nose platform,}}}}
\ {\bibfield  {journal} {\bibinfo  {journal} {Biomaterials}\ }\textbf {\bibinfo
  {volume} {33}},\ \bibinfo {pages} {1722} (\bibinfo {year}
  {2012}{\natexlab{a}})}\BibitemShut {NoStop}%
\bibitem [{\citenamefont {Farahi}\ \emph {et~al.}(2012)\citenamefont {Farahi},
  \citenamefont {Passian}, \citenamefont {Tetard},\ and\ \citenamefont
  {Thundat}}]{farahi_critical_2012}%
  \BibitemOpen
  \bibfield  {author} {\bibinfo {author} {\bibfnamefont {R.~H.}\ \bibnamefont
  {Farahi}}, \bibinfo {author} {\bibfnamefont {A.}~\bibnamefont {Passian}},
  \bibinfo {author} {\bibfnamefont {L.}~\bibnamefont {Tetard}}, \ and\ \bibinfo
  {author} {\bibfnamefont {T.}~\bibnamefont {Thundat}},\ }\href@noop {} %{{\selectlanguage {english}\emph {\bibinfo {title}  {Critical Issues in Sensor Science To Aid Food and Water Safety,}}}}
\ {\bibfield  {journal} {\bibinfo
  {journal} {{ACS} Nano}\ }\textbf {\bibinfo {volume} {6}},\ \bibinfo {pages}
  {4548} (\bibinfo {year} {2012})}\BibitemShut {NoStop}%
\bibitem [{\citenamefont {Silverman}(2002)}]{silverman_organic_2002}%
  \BibitemOpen
  \bibfield  {author} {\bibinfo {author} {\bibfnamefont {R.~B.}\ \bibnamefont
  {Silverman}},\ }\href@noop {} {{\selectlanguage {english}\emph {\bibinfo
  {title} {The Organic Chemistry of Enzyme-catalyzed Reactions}}}}\ (\bibinfo
  {publisher} {Academic Press, London},\ \bibinfo {year} {2002})\BibitemShut {NoStop}%
\bibitem [{\citenamefont {Haffenden}\ \emph {et~al.}(2001)\citenamefont
  {Haffenden}, \citenamefont {Yaylayan},\ and\ \citenamefont
  {Fortin}}]{haffenden_investigation_2001}%
  \BibitemOpen
  \bibfield  {author} {\bibinfo {author} {\bibfnamefont {L.}~\bibnamefont
  {Haffenden}}, \bibinfo {author} {\bibfnamefont {V.}~\bibnamefont {Yaylayan}},
  \ and\ \bibinfo {author} {\bibfnamefont {J.}~\bibnamefont {Fortin}},\ }\href@noop {} %{{\selectlanguage {english}\emph {\bibinfo {title}  {Investigation of vibrational theory of olfaction with variously labelled benzaldehydes,}}}}
\ {\bibfield  {journal} {\bibinfo  {journal} {Food Chem.}\ }\textbf {\bibinfo
  {volume} {73}},\ \bibinfo {pages} {67} (\bibinfo {year} {2001})}\BibitemShut
  {NoStop}%
\bibitem [{\citenamefont {Dyson}(1938)}]{dyson_scientific_1938}%
  \BibitemOpen
  \bibfield  {author} {\bibinfo {author} {\bibfnamefont {G.~M.}\ \bibnamefont
  {Dyson}},\ }\href@noop {} %{{\selectlanguage {english}\emph {\bibinfo {title}  {The scientific basis of odour,}}}}
\ {\bibfield  {journal} {\bibinfo  {journal} {Chem.
  Ind.}\ }\textbf {\bibinfo {volume} {57}},\ \bibinfo {pages} {647} (\bibinfo
  {year} {1938})}\BibitemShut {NoStop}%
\bibitem [{\citenamefont {Wright}(1977)}]{wright_odor_1977}%
  \BibitemOpen
  \bibfield  {author} {\bibinfo {author} {\bibfnamefont {R.~H.}\ \bibnamefont
  {Wright}},\ }\href@noop {} %{{\selectlanguage {english}\emph {\bibinfo {title}  {Odor and molecular vibration: neural coding of olfactory information,}}}}
\ {\bibfield  {journal} {\bibinfo  {journal} {J.
  Theor. Biol.}\ }\textbf {\bibinfo {volume} {64}},\ \bibinfo {pages} {473}
  (\bibinfo {year} {1977})}\BibitemShut {NoStop}%
\bibitem [{\citenamefont {Wright}(1982)}]{wright_sense_1982}%
  \BibitemOpen
  \bibfield  {author} {\bibinfo {author} {\bibfnamefont {R.~H.}\ \bibnamefont
  {Wright}},\ }\href@noop {} {{\selectlanguage {english}\emph {\bibinfo {title}
  {The sense of smell}}}}\ (\bibinfo  {publisher} {{CRC} Press},\ \bibinfo {year} {1982})\BibitemShut {NoStop}%
\bibitem [{\citenamefont {Lambe}\ and\ \citenamefont
  {Jaklevic}(1968)}]{lambe_molecular_1968}%
  \BibitemOpen
  \bibfield  {author} {\bibinfo {author} {\bibfnamefont {J.}~\bibnamefont
  {Lambe}}\ and\ \bibinfo {author} {\bibfnamefont {R.~C.}\ \bibnamefont
  {Jaklevic}},\ }\href@noop {} %{{\selectlanguage {english}\emph {\bibinfo {title}  {Molecular Vibration Spectra by Inelastic Electron Tunneling,}}}}
\ {\bibfield  {journal} {\bibinfo  {journal} {Phys. Rev.}\ }\textbf {\bibinfo
  {volume} {165}},\ \bibinfo {pages} {821} (\bibinfo {year}
  {1968})}\BibitemShut {NoStop}%
\bibitem [{\citenamefont {Adkins}\ and\ \citenamefont
  {Phillips}(1985)}]{adkins_inelastic_1985}%
  \BibitemOpen
  \bibfield  {author} {\bibinfo {author} {\bibfnamefont {C.~J.}\ \bibnamefont
  {Adkins}}\ and\ \bibinfo {author} {\bibfnamefont {W.~A.}\ \bibnamefont
  {Phillips}},\ }\href@noop {} %{{\selectlanguage {english}\emph {\bibinfo {title}  {Inelastic electron tunnelling spectroscopy,}}}}
\ {\bibfield  {journal} {\bibinfo  {journal} {J. Phys. C}\ }\textbf {\bibinfo
  {volume} {18}},\ \bibinfo {pages} {1313} (\bibinfo {year}
  {1985})}\BibitemShut {NoStop}%
\bibitem [{\citenamefont {Marcus}(1956)}]{marcus_theory_1956}%
  \BibitemOpen
  \bibfield  {author} {\bibinfo {author} {\bibfnamefont {R.~A.}\ \bibnamefont
  {Marcus}},\ }\href@noop {} %{{\selectlanguage {english}\emph {\bibinfo {title}  {On the Theory of Oxidation?Reduction Reactions Involving Electron Transfer. I,}}}}
\ {\bibfield  {journal} {\bibinfo  {journal} {J. Chem. Phys.}\ }\textbf
  {\bibinfo {volume} {24}},\ \bibinfo {pages} {966} (\bibinfo {year}
  {1956})}\BibitemShut {NoStop}%
\bibitem [{\citenamefont {Marcus}(1964)}]{marcus_chemical_1964}%
  \BibitemOpen
  \bibfield  {author} {\bibinfo {author} {\bibfnamefont {R.~A.}\ \bibnamefont
  {Marcus}},\ }\href@noop {} %{{\selectlanguage {english}\emph {\bibinfo {title}  {Chemical and Electrochemical Electron-Transfer Theory,}}}}
\ {\bibfield  {journal} {\bibinfo  {journal} {Annu. Rev. Phys. Chem.}\ }\textbf
  {\bibinfo {volume} {15}},\ \bibinfo {pages} {155} (\bibinfo {year}
  {1964})}\BibitemShut {NoStop}%
\bibitem [{\citenamefont {Marcus}(1965)}]{marcus_theory_1965}%
  \BibitemOpen
  \bibfield  {author} {\bibinfo {author} {\bibfnamefont {R.~A.}\ \bibnamefont
  {Marcus}},\ }\href@noop {} %{{\selectlanguage {english}\emph {\bibinfo {title}  {On the Theory of Electron-Transfer Reactions. VI. Unified Treatment for Homogeneous and Electrode Reactions,}}}}
\ {\bibfield  {journal} {\bibinfo  {journal} {J. Chem. Phys.}\ }\textbf
  {\bibinfo {volume} {43}},\ \bibinfo {pages} {679} (\bibinfo {year}
  {1965})}\BibitemShut {NoStop}%
\bibitem [{\citenamefont {Marcus}\ and\ \citenamefont
  {Sutin}(1985)}]{marcus_electron_1985}%
  \BibitemOpen
  \bibfield  {author} {\bibinfo {author} {\bibfnamefont {R.}~\bibnamefont
  {Marcus}}\ and\ \bibinfo {author} {\bibfnamefont {N.}~\bibnamefont {Sutin}},\
  }\href@noop {} %{{\selectlanguage {english}\emph {\bibinfo {title}  {Electron transfers in chemistry and biology,}}}}
\ {\bibfield  {journal} {\bibinfo  {journal} {Biochim. Biophys. Acta.}\
  }\textbf {\bibinfo {volume} {811}},\ \bibinfo {pages} {265} (\bibinfo {year}
  {1985})}\BibitemShut {NoStop}%
\bibitem [{\citenamefont {Bixon}\ and\ \citenamefont
  {Jortner}(1999)}]{bixon_electron_1999}%
  \BibitemOpen
  \bibfield  {author} {\bibinfo {author} {\bibfnamefont {M.}~\bibnamefont
  {Bixon}}\ and\ \bibinfo {author} {\bibfnamefont {J.}~\bibnamefont
  {Jortner}},\ }\href@noop {} %{{\selectlanguage {english}\emph {\bibinfo {title}  {Electron Transfer. From Isolated Molecules to Biomolecules,}}}}
\ {\bibfield  {journal} {\bibinfo  {journal} {Adv.
  Chem. Phys.}\ }\textbf {\bibinfo {volume} {106}},\ \bibinfo {pages} {35}
  (\bibinfo {year} {1999})}\BibitemShut {NoStop}%
\bibitem [{\citenamefont
  {Holstein}(1959{\natexlab{a}})}]{holstein_studies_1959-1}%
  \BibitemOpen
  \bibfield  {author} {\bibinfo {author} {\bibfnamefont {T.}~\bibnamefont
  {Holstein}},\ }\href@noop {} %{{\selectlanguage {english}\emph {\bibinfo {title}  {Studies of polaron motion: Part I. The molecular-crystal model,}}}}
\ {\bibfield  {journal} {\bibinfo  {journal} {Ann. Phys.}\ }\textbf {\bibinfo
  {volume} {8}},\ \bibinfo {pages} {325} (\bibinfo {year}
  {1959}{\natexlab{a}})}\BibitemShut {NoStop}%
\bibitem [{\citenamefont
  {Holstein}(1959{\natexlab{b}})}]{holstein_studies_1959}%
  \BibitemOpen
  \bibfield  {author} {\bibinfo {author} {\bibfnamefont {T.}~\bibnamefont
  {Holstein}},\ }\href@noop {} %{{\selectlanguage {english}\emph {\bibinfo {title}  {Studies of polaron motion: Part II. The ``small'' polaron,}}}}
\ {\bibfield  {journal} {\bibinfo  {journal} {Ann. Phys.}\ }\textbf {\bibinfo
  {volume} {8}},\ \bibinfo {pages} {343} (\bibinfo {year}
  {1959}{\natexlab{b}})}\BibitemShut {NoStop}%
\bibitem [{\citenamefont {Jackson}\ and\ \citenamefont
  {Silbey}(1983)}]{JacksonSilbey1983}%
  \BibitemOpen
  \bibfield  {author} {\bibinfo {author} {\bibfnamefont {B.}~\bibnamefont
  {Jackson}}\ and\ \bibinfo {author} {\bibfnamefont {R.}~\bibnamefont
  {Silbey}},\ }\href@noop {} %{{\selectlanguage {english}\emph {\bibinfo {title}  {On the calculation of transfer rates between impurity states in solids,}}}}
\ {\bibfield  {journal} {\bibinfo  {journal} {J.
  Chem. Phys.}\ }\textbf {\bibinfo {volume} {78}},\ \bibinfo {pages} {4193}
  (\bibinfo {year} {1983})}\BibitemShut {NoStop}%
\bibitem [{\citenamefont {Nazir}(2009)}]{nazir09}%
  \BibitemOpen
  \bibfield  {author} {\bibinfo {author} {\bibfnamefont {A.}~\bibnamefont
  {Nazir}},\ }\href@noop {} %{{\selectlanguage {english}\emph {\bibinfo {title}  {Correlation-Dependent Coherent to Incoherent Transitions in Resonant Energy Transfer Dynamics,}}}}
\ {\bibfield  {journal} {\bibinfo  {journal} {Phys.
  Rev. Lett.}\ }\textbf {\bibinfo {volume} {103}},\ \bibinfo {pages} {146404}
  (\bibinfo {year} {2009})}\BibitemShut {NoStop}%
\bibitem [{\citenamefont {Jang}\ \emph {et~al.}(2008)\citenamefont {Jang},
  \citenamefont {Cheng}, \citenamefont {Reichman},\ and\ \citenamefont
  {Eaves}}]{jang_theory_2008}%
  \BibitemOpen
  \bibfield  {author} {\bibinfo {author} {\bibfnamefont {S.}~\bibnamefont
  {Jang}}, \bibinfo {author} {\bibfnamefont {Y.-C.}\ \bibnamefont {Cheng}},
  \bibinfo {author} {\bibfnamefont {D.~R.}\ \bibnamefont {Reichman}}, \ and\
  \bibinfo {author} {\bibfnamefont {J.~D.}\ \bibnamefont {Eaves}},\ }\href@noop {} %{{\selectlanguage {english}\emph {\bibinfo {title}  {Theory of coherent resonance energy transfer,}}}}
\ {\bibfield  {journal} {\bibinfo  {journal} {J. Chem. Phys.}\ }\textbf
  {\bibinfo {volume} {129}},\ \bibinfo {pages} {101104} (\bibinfo {year}
  {2008})}\BibitemShut {NoStop}%
\bibitem [{\citenamefont {Jang}(2009)}]{jang09}%
  \BibitemOpen
  \bibfield  {author} {\bibinfo {author} {\bibfnamefont {S.}~\bibnamefont
  {Jang}},\ }\href@noop {} %{{\selectlanguage {english}\emph {\bibinfo {title}  {Theory of coherent resonance energy transfer for coherent initial condition,}}}}
\ {\bibfield  {journal} {\bibinfo  {journal} {J. Chem.
  Phys.}\ }\textbf {\bibinfo {volume} {131}},\ \bibinfo {pages} {164101}
  (\bibinfo {year} {2009})}\BibitemShut {NoStop}%
\bibitem [{\citenamefont {Jang}(2011)}]{Jang2011}%
  \BibitemOpen
  \bibfield  {author} {\bibinfo {author} {\bibfnamefont {S.}~\bibnamefont
  {Jang}},\ }\href@noop {} %{{\selectlanguage {english}\emph {\bibinfo {title}  {Theory of multichromophoric coherent resonance energy transfer: A polaronic quantum master equation approach,}}}}
\ {\bibfield  {journal} {\bibinfo  {journal} {J. Chem.
  Phys.}\ }\textbf {\bibinfo {volume} {135}},\ \bibinfo {eid} {034105}
  (\bibinfo {year} {2011})}\BibitemShut {NoStop}%
\bibitem [{\citenamefont {McCutcheon}\ and\ \citenamefont
  {Nazir}(2010)}]{McCutcheonNazir2010}%
  \BibitemOpen
  \bibfield  {author} {\bibinfo {author} {\bibfnamefont {D.~P.~S.}\
  \bibnamefont {McCutcheon}}\ and\ \bibinfo {author} {\bibfnamefont
  {A.}~\bibnamefont {Nazir}},\ }\href@noop {} %{{\selectlanguage {english}\emph {\bibinfo {title}  {Quantum dot Rabi rotations beyond the weak exciton-phonon coupling regime,}}}}
\ {\bibfield  {journal} {\bibinfo
  {journal} {New J. Phys.}\ }\textbf {\bibinfo {volume} {12}},\ \bibinfo
  {pages} {113042} (\bibinfo {year} {2010})}\BibitemShut {NoStop}%
\bibitem [{\citenamefont {McCutcheon}\ and\ \citenamefont
  {Nazir}(2011{\natexlab{a}})}]{mccutcheon11}%
  \BibitemOpen
  \bibfield  {author} {\bibinfo {author} {\bibfnamefont {D.~P.~S.}\
  \bibnamefont {McCutcheon}}\ and\ \bibinfo {author} {\bibfnamefont
  {A.}~\bibnamefont {Nazir}},\ }\href@noop {} %{{\selectlanguage {english}\emph {\bibinfo {title}  {Coherent and incoherent dynamics in excitonic energy transfer: Correlated fluctuations and off-resonance effects,}}}}
\ {\bibfield  {journal} {\bibinfo
  {journal} {Phys. Rev. B}\ }\textbf {\bibinfo {volume} {83}},\ \bibinfo
  {pages} {165101} (\bibinfo {year} {2011}{\natexlab{a}})}\BibitemShut
  {NoStop}%
\bibitem [{\citenamefont {Kolli}\ \emph {et~al.}(2011)\citenamefont {Kolli},
  \citenamefont {Nazir},\ and\ \citenamefont {Olaya-Castro}}]{KolliNazir2011}%
  \BibitemOpen
  \bibfield  {author} {\bibinfo {author} {\bibfnamefont {A.}~\bibnamefont
  {Kolli}}, \bibinfo {author} {\bibfnamefont {A.}~\bibnamefont {Nazir}}, \ and\
  \bibinfo {author} {\bibfnamefont {A.}~\bibnamefont {Olaya-Castro}},\
  }\href@noop {} %{{\selectlanguage {english}\emph {\bibinfo {title}  {Electronic excitation dynamics in multichromophoric systems described via a polaron-representation master equation,}}}}
\ {\bibfield  {journal} {\bibinfo  {journal} {J. Chem. Phys.}\
  }\textbf {\bibinfo {volume} {135}} \ \bibinfo
  {pages} {154112} (\bibinfo {year} {2011})}\BibitemShut
  {NoStop}%
\bibitem [{\citenamefont {McCutcheon}\ and\ \citenamefont
  {Nazir}(2011{\natexlab{b}})}]{mccutcheon11b}%
  \BibitemOpen
  \bibfield  {author} {\bibinfo {author} {\bibfnamefont {D.~P.~S.}\
  \bibnamefont {McCutcheon}}\ and\ \bibinfo {author} {\bibfnamefont
  {A.}~\bibnamefont {Nazir}},\ }\href@noop {} %{{\selectlanguage {english}\emph {\bibinfo {title}  {Consistent treatment of coherent and incoherent energy transfer dynamics using a variational master equation,}}}}
\ {\bibfield  {journal} {\bibinfo
  {journal} {J. Chem. Phys.}\ }\textbf {\bibinfo {volume} {135}},\ \bibinfo
  {pages} {114501} (\bibinfo {year} {2011}{\natexlab{b}})}\BibitemShut
  {NoStop}%
\bibitem [{\citenamefont {Buck}\ and\ \citenamefont
  {Axel}(1991)}]{buck_novel_1991}%
  \BibitemOpen
  \bibfield  {author} {\bibinfo {author} {\bibfnamefont {L.}~\bibnamefont
  {Buck}}\ and\ \bibinfo {author} {\bibfnamefont {R.}~\bibnamefont {Axel}},\
  }\href@noop {} %{{\selectlanguage {english}\emph {\bibinfo {title}  {A novel multigene family may encode odorant receptors: A molecular basis for odor recognition,}}}}
\ {\bibfield  {journal} {\bibinfo  {journal} {Cell}\ }\textbf {\bibinfo
  {volume} {65}},\ \bibinfo {pages} {175} (\bibinfo {year} {1991})}\BibitemShut
  {NoStop}%
%\bibitem [{\citenamefont {Wang}\ \emph {et~al.}(2003)\citenamefont {Wang},
%  \citenamefont {Luthey-Schulten},\ and\ \citenamefont
%  {Suslick}}]{wang_is_2003}%
%  \BibitemOpen
%  \bibfield  {author} {\bibinfo {author} {\bibfnamefont {J.}~\bibnamefont
%  {Wang}}, \bibinfo {author} {\bibfnamefont {Z.~A.}\ \bibnamefont
%  {Luthey-Schulten}}, \ and\ \bibinfo {author} {\bibfnamefont {K.~S.}\
%  \bibnamefont {Suslick}},\ }\href@noop {} %{{\selectlanguage {english}\emph {\bibinfo {title}  {Is the olfactory receptor a metalloprotein?,}}}}
%\ {\bibfield  {journal} {\bibinfo  {journal} {PNAS}\ }\textbf {\bibinfo
%  {volume} {100}},\ \bibinfo {pages} {3035} (\bibinfo {year}
%  {2003})}\BibitemShut {NoStop}%
%\bibitem [{\citenamefont {Duan}\ \emph {et~al.}(2012)\citenamefont {Duan},
%  \citenamefont {Block}, \citenamefont {Li}, \citenamefont {Connelly},
%  \citenamefont {Zhang}, \citenamefont {Huang}, \citenamefont {Su},
%  \citenamefont {Pan}, \citenamefont {Wu}, \citenamefont {Chi}, \citenamefont
%  {Thomas}, \citenamefont {Zhang}, \citenamefont {Ma}, \citenamefont
%  {Matsunami}, \citenamefont {Chen},\ and\ \citenamefont
%  {Zhuang}}]{duan_crucial_2012}%
%  \BibitemOpen
%  \bibfield  {author} {\bibinfo {author} {\bibfnamefont {X.}~\bibnamefont
%  {Duan}}, \bibinfo {author} {\bibfnamefont {E.}~\bibnamefont {Block}},
%  \bibinfo {author} {\bibfnamefont {Z.}~\bibnamefont {Li}}, \bibinfo {author}
%  {\bibfnamefont {T.}~\bibnamefont {Connelly}}, \bibinfo {author}
%  {\bibfnamefont {J.}~\bibnamefont {Zhang}}, \bibinfo {author} {\bibfnamefont
%  {Z.}~\bibnamefont {Huang}}, \bibinfo {author} {\bibfnamefont
%  {X.}~\bibnamefont {Su}}, \bibinfo {author} {\bibfnamefont {Y.}~\bibnamefont
%  {Pan}}, \bibinfo {author} {\bibfnamefont {L.}~\bibnamefont {Wu}}, \bibinfo
%  {author} {\bibfnamefont {Q.}~\bibnamefont {Chi}}, \bibinfo {author}
%  {\bibfnamefont {S.}~\bibnamefont {Thomas}}, \bibinfo {author} {\bibfnamefont
%  {S.}~\bibnamefont {Zhang}}, \bibinfo {author} {\bibfnamefont
%  {M.}~\bibnamefont {Ma}}, \bibinfo {author} {\bibfnamefont {H.}~\bibnamefont
%  {Matsunami}}, \bibinfo {author} {\bibfnamefont {G.-Q.}\ \bibnamefont {Chen}},
%  \ and\ \bibinfo {author} {\bibfnamefont {H.}~\bibnamefont {Zhuang}},\ }\href@noop {} %{{\selectlanguage {english}\emph {\bibinfo {title}  {Crucial role of copper in detection of metal-coordinating odorants,}}}}
%\ {\bibfield  {journal} {\bibinfo
%  {journal} {PNAS}\ }\textbf {\bibinfo {volume} {109}},\ \bibinfo {pages}
%  {3492} (\bibinfo {year} {2012})}\BibitemShut {NoStop}%
%\bibitem [{Note1()}]{Note1}%
%  \BibitemOpen
%  \bibinfo {note} {Supplemental Material.}\BibitemShut {Stop}% 
\bibitem [{Note2()}]{Note2}%
  \BibitemOpen
  \bibinfo {note} {For $\omega _0=200$~meV and $T=300$~K this corresponds to
  the odorant initial state being essentially the ground state, and is
  therefore consistent with the initial condition in the rate
  analysis.}\BibitemShut {Stop}%
\bibitem [{Note3()}]{Note3}%
  \BibitemOpen
  \bibinfo {note} {The DA dynamics is generally well fitted by a biexponential
  form.}\BibitemShut {Stop}%
\bibitem [{Note4()}]{Note4}%
  \BibitemOpen
  \bibinfo {note} {Note that the ET rates used to calculate the dashed curves
  in panels (i - iii) of Fig.~\ref {rates-dynamics} do agree with the MJ
  rates.}\BibitemShut {Stop}% 
\bibitem [{\citenamefont {Remus}\ and\ \citenamefont
  {Blencowe}(2012)}]{remus_damping_2012}%
  \BibitemOpen
  \bibfield  {author} {\bibinfo {author} {\bibfnamefont {L.~G.}\ \bibnamefont
  {Remus}}\ and\ \bibinfo {author} {\bibfnamefont {M.~P.}\ \bibnamefont
  {Blencowe}},\ }\href@noop {} %{{\selectlanguage {english}\emph {\bibinfo {title}  {Damping and decoherence of Fock states in a nanomechanical resonator due to two-level systems,}}}}
\ {\bibfield  {journal} {\bibinfo  {journal} {Phys. Rev. B}\ }\textbf {\bibinfo
  {volume} {86}},\ \bibinfo {pages} {205419} (\bibinfo {year}
  {2012})}\BibitemShut {NoStop}%
\bibitem [{\citenamefont {Lambert}\ and\ \citenamefont
  {Nori}(2008)}]{lambert08}%
  \BibitemOpen
  \bibfield  {author} {\bibinfo {author} {\bibfnamefont {N.}~\bibnamefont
  {Lambert}}\ and\ \bibinfo {author} {\bibfnamefont {F.}~\bibnamefont {Nori}},\
  }\href@noop {} %{{\selectlanguage {english}\emph {\bibinfo {title}  {Detecting quantum-coherent nanomechanical oscillations using the current-noise spectrum of a double quantum dot,}}}}
\ {\bibfield  {journal} {\bibinfo  {journal} {Phys. Rev. B}\
  }\textbf {\bibinfo {volume} {78}},\ \bibinfo {pages} {214302} (\bibinfo
  {year} {2008})}\BibitemShut {NoStop}%
\bibitem [{\citenamefont {Armour}\ and\ \citenamefont
  {Blencowe}(2008{\natexlab{a}})}]{armour_probing_2008}%
  \BibitemOpen
  \bibfield  {author} {\bibinfo {author} {\bibfnamefont {A.~D.}\ \bibnamefont
  {Armour}}\ and\ \bibinfo {author} {\bibfnamefont {M.~P.}\ \bibnamefont
  {Blencowe}},\ }\href@noop {} %{{\selectlanguage {english}\emph {\bibinfo {title}  {Probing the quantum coherence of a nanomechanical resonator using a superconducting qubit: I. Echo scheme,}}}}
\ {\bibfield  {journal} {\bibinfo  {journal} {New J. Phys.}\ }\textbf {\bibinfo
  {volume} {10}},\ \bibinfo {pages} {095004} (\bibinfo {year}
  {2008}{\natexlab{a}})}
  \BibitemShut {NoStop}%
\bibitem [{\citenamefont {Armour}\ and\ \citenamefont
  {Blencowe}(2008{\natexlab{b}})}]{armour_probing_2008b}%
  \BibitemOpen
  \bibfield  {author} {\bibinfo {author} {\bibfnamefont {A.~D.}\ \bibnamefont
  {Armour}}\ and\ \bibinfo {author} {\bibfnamefont {M.~P.}\ \bibnamefont
  {Blencowe}},\ }\href@noop {} %{{\selectlanguage {english}\emph {\bibinfo {title}  {Probing the quantum coherence of a nanomechanical resonator using a superconducting qubit: II. Implementation,}}}}
\ {\bibfield  {journal} {\bibinfo  {journal} {New
  J. Phys.}\ }\textbf {\bibinfo {volume} {10}},\ \bibinfo {pages} {095005}
  (\bibinfo {year} {2008}{\natexlab{b}})}\BibitemShut {NoStop}%
\bibitem [{\citenamefont {Xu}\ and\ \citenamefont {Schulten}(1994)}]{xu94}%
  \BibitemOpen
  \bibfield  {author} {\bibinfo {author} {\bibfnamefont {D.}~\bibnamefont
  {Xu}}\ and\ \bibinfo {author} {\bibfnamefont {K.}~\bibnamefont {Schulten}},\
  }\href@noop {} %{{\selectlanguage {english}\emph {\bibinfo {title}  {Coupling of protein motion to electron transfer in a photosynthetic reaction center: investigating the low temperature behavior in the framework of the spin-boson model,}}}}
\ {\bibfield  {journal} {\bibinfo  {journal} {Chem. Phys.}\
  }\textbf {\bibinfo {volume} {182}},\ \bibinfo {pages} {91} (\bibinfo {year}
  {1994})}\BibitemShut {NoStop}%
\bibitem [{\citenamefont {Blankenship}(2002)}]{blankenship_molecular_2002}%
  \BibitemOpen
  \bibfield  {author} {\bibinfo {author} {\bibfnamefont {R.~E.}\ \bibnamefont
  {Blankenship}},\ }\href@noop {} {{\selectlanguage {english}\emph {\bibinfo
  {title} {Molecular mechanisms of photosynthesis}}}}\ (\bibinfo  {publisher}
  {Blackwell},\ \bibinfo {address} {Oxford},\ \bibinfo
  {year} {2002})\BibitemShut {NoStop}%
\bibitem [{\citenamefont {Lee}\ \emph {et~al.}(2012{\natexlab{b}})\citenamefont
  {Lee}, \citenamefont {Jin}, \citenamefont {Song}, \citenamefont {Hong},\ and\
  \citenamefont {Park}}]{lee_bioelectronic_2012}%
  \BibitemOpen
  \bibfield  {author} {\bibinfo {author} {\bibfnamefont {S.~H.}\ \bibnamefont
  {Lee}}, \bibinfo {author} {\bibfnamefont {H.~J.}\ \bibnamefont {Jin}},
  \bibinfo {author} {\bibfnamefont {H.~S.}\ \bibnamefont {Song}}, \bibinfo
  {author} {\bibfnamefont {S.}~\bibnamefont {Hong}}, \ and\ \bibinfo {author}
  {\bibfnamefont {T.~H.}\ \bibnamefont {Park}},\ }\href@noop {} %{{\selectlanguage {english}\emph {\bibinfo {title}  {Bioelectronic nose with high sensitivity and selectivity using chemically functionalized carbon nanotube combined with human olfactory receptor,}}}}
\ {\bibfield  {journal} {\bibinfo  {journal} {J. Biotechnol.}\ }\textbf
  {\bibinfo {volume} {157}},\ \bibinfo {pages} {467} (\bibinfo {year}
  {2012}{\natexlab{b}})}\BibitemShut {NoStop}%
\bibitem [{\citenamefont {Du}\ \emph {et~al.}(2013)\citenamefont {Du},
  \citenamefont {Wu}, \citenamefont {Liu}, \citenamefont {Huang},\ and\
  \citenamefont {Wang}}]{du_recent_2013}%
  \BibitemOpen
  \bibfield  {author} {\bibinfo {author} {\bibfnamefont {L.}~\bibnamefont
  {Du}}, \bibinfo {author} {\bibfnamefont {C.}~\bibnamefont {Wu}}, \bibinfo
  {author} {\bibfnamefont {Q.}~\bibnamefont {Liu}}, \bibinfo {author}
  {\bibfnamefont {L.}~\bibnamefont {Huang}}, \ and\ \bibinfo {author}
  {\bibfnamefont {P.}~\bibnamefont {Wang}},\ }\href@noop {} %{{\selectlanguage {english}\emph {\bibinfo {title}  {Recent advances in olfactory receptor-based biosensors,}}}}
\ {\bibfield  {journal} {\bibinfo  {journal} {Biosens. Bioelectron.}\ }\textbf
  {\bibinfo {volume} {42}},\ \bibinfo {pages} {570} (\bibinfo {year}
  {2013})}\BibitemShut {NoStop}%
  %
\end{thebibliography}
\end{document}